\documentclass[english,pra,showpacs,twocolumn,amsmath]{revtex4}
\usepackage{graphicx}
\usepackage{simplewick}

\begin{document}

\title{A dense Bose fluid at zero temperature: condensation and clusters in liquid $^4$He }

\author{V I Kruglov$^{*}$ and M J Collett}

\affiliation{Physics Department, The University of Auckland, Private Bag 92019, Auckland, New Zealand}

\affiliation{$^{*}$Corresponding author: v.kruglov@auckland.ac.nz}

\begin{abstract}
We present a full set of wave equations describing a dense Bose fluid, applicable both to non-ideal gases and to liquid $^4$He. The phonon spectrum in liquid $^4$He is found and the fraction of condensed particles is calculated at zero temperature for a wide range of densities. The theory also yields the ground-state energy for the quantum liquid $^4$He in agreement to high accuracy with Monte Carlo simulations and experimental data at low pressure. We also present the derivation of a generalized Hartree-Fock equation describing roton clusters in low temperature liquid $^4$He, allowing us to confirm that, at low enough temperatures and for a wide range of pressures, the stable clusters consist of 13 bound atoms. 
\end{abstract}


\maketitle

\section{Introduction}

The study of the structural and dynamical properties of liquid helium at low temperatures (below the lambda-transition) has a long history. The analogy between liquid $^4$He and an ideal Bose-Einstein gas was first suggested by London \cite{Lo,Lon}. The fact that no lambda-transition has been found in $^3$He supports this viewpoint. After London's suggestion, Tisza \cite{Tis} showed that the analogy between liquid $^4$He and an ideal Bose-Einstein gas is useful in understanding the transport properties of He II. Tisza realized that the presence of the condensed fraction would make necessary a two-fluid hydrodynamical description for liquid helium; his idea that a two-fluid description can also be applied to He II has been verified by many experimental works \cite{Pesh,Andr,FLon}.

 However, Tisza's model did not appear to be completely correct, and the two-fluid model was subsequently developed by Landau \cite{Lan} with no reference to Bose-Einstein condensation (BEC). Modern understanding of superfluidity is also based on the quantization of circulation in a superfluid, which was first proposed by Onsager and was used independently by Feynman in his study of the critical velocity for superfluidity \cite{Fe,Khal}.
Landau also modified the dispersion relation for the rotons \cite{Land} in order to obtain a better agreement with thermodynamic properties. Feynman proposed a relation between the energy spectrum and the structure factor\cite{RF,CoF} that verifies Landau's dispersion relation. This spectrum was also measured in neutron scattering experiments with great accuracy by several groups, in particular by Henshaw and Woods \cite{Hen}. 
 
Support for London's viewpoint comes from later theoretical work \cite{Bog,Mat,Feyn,Ches} which shows how a system of interacting particles can exhibit a lambda-transition corresponding to the ideal gas transition. However a further complication arises that the theory of superfluidity \cite{Lan,Fey} is apparently independent of the ideal gas analogy. Some other works, for an example \cite{Bog,Pen} do suggest the ideal gas analogy, but there are no strict experimental and theoretical confirmations of such a simple analogy for liquid $^4$He,
only for a dilute gas \cite{Pit}. 

We note that Feynman also proposed \cite{Fe} a model of the roton as a localized vortex ring with a characteristic size of the order of the mean atomic distance in liquid $^4$He. Vortex rings were observed experimentally; however, there has been no experimental confirmation that these vortex rings are indeed the roton excitations.

The more recent realization of Bose-condensed gases with alkali elements \cite{And,Davis} provided a great opportunity to test a new regime of matter that until then was considered a theoretical concept. The theoretical basis for the description of these systems is presented in Ref. \cite{Dalfovo}. 

A number of methods have been also suggested for applications to quantum fluids. For an example, the phenomenological density-functional for liquid $^4$He \cite{Dup} has proved to be quite reliable for helium films \cite{Pav,Cheng}, vortices in bulk liquid \cite{Da}, droplets \cite{Dall}, and structural and dynamical properties of superfluid helium \cite{Dal}. For a review of quantum fluid theories also see Refs. \cite{Noz,Gru,Ap,Bau,Bal,Lip}. The renormalization group method was used recently \cite{Kaup,Cam,Pog} to describe the heat capacity in liquid $^4$He.

We review in Sec.~II the Gross-Pitaevskii (GP) equation for a nearly ideal Bose gas, obtained by the Hartree-Fock approach.
In Sec.~III we derive a system of two coupled wave equations which can describe either a dense Bose gas or liquid $^4$He. The results for the nearly ideal Bose gas follow from these as a limiting case. As a sample application we find the solution for the condensate wave function of superfluid $^4$He propagating in a channel. General results for the ground state energy and the phonon spectrum are derived in Sec.~IV. We note that the form of this spectrum is similar to that of Bogoliubov \cite{Bog}, but unlike the Bogoliubov spectrum its validity is not restricted to small densities. 

Based on the results of Sec.~IV we develop in Sec.~V the theory of Bose-Einstein condensation of liquid $^4$He at zero temperature for a wide range of densities and pressures. When the mass density is $\rho=0.145~{\rm g~cm}^{-3}$ we find that the condensate and excitation fractions are $\rho_c/\rho=0.528$ and $\rho_{ex}/\rho=0.472$ respectively. 
This is significantly different from the crude estimate one may find in many works (e.g.\ \cite{Pen}) that roughly $10\%$ of the atoms are `condensed' in liquid $^4$He at low pressure and zero temperature. 
Our calculations of the condensate and excitation fractions are based on the equation of state and the equation for the energy of the ground state per particle at zero temperature, and are confirmed by Monte Carlo simulations and the experimental data. 

In Sec.~VI we find the ground state energy of liquid $^4$He based on an effective Hamiltonian, derived from the approximate Heisenberg equation found in Sec.~III. Diagonalization of the effective Hamiltonian allows us to calculate the excitation and condensed densities. 
An important step is the introduction of a cut-off in momentum space at the speed of the second sound. 
The condensate and excitation fractions found by this method are in remarkable agreement with the ones found in Sec.~V 
(and hence also with Monte Carlo simulations and experimental data). 
The theory also yields the ground state energy per particle for liquid $^4$He at zero pressure and temperature as ${\cal E}_0/k_B=-7.12~{\rm K}$, which is very close to the value obtained by Monte Carlo simulations.  
 
In Sec.~VII a generalized Hartree-Fock (GHF) equation is derived that describes the ground state of a roton cluster in a quantum liquid in the mean field approximation. In particular, it allows us (see Appendix D) to evaluate the number of bound atoms in such a cluster for a wide range of pressures. 
At low temperatures and typical pressures the stable clusters in liquid helium  consist of $13$ bound helium atoms \cite{Kr,Kru}, presumably in the form of a central atom surrounded by an icosahedral shell of $12$ atoms. 
The number of atoms in a cluster can however be less than $13$ if the pressure is negative. 


Some important supplementary results are presented in Appendices A-D.

\section{Hartree-Fock and GP approximations}

The theory of condensation of a nearly ideal Bose-Einstein gas is presented, for example, in \cite{Pit,Peth,Griff}. We introduce in this section a Gross-Pitaevskii (GP) approximation that does not require the explicit replacement of the true interatomic potential by a delta-function pseudo-potential. Our approach is based on a modified Born approximation (MBA) devised in \cite{Kru} and reproduced in Appendix A. The GP theory using the Hartree-Fock approach and the MBA may be viewed as the limiting case of the theory for a dense Bose gas and liquid $^4$He developed in the following sections. 

It has been emphasized \cite{Pro} that the Hamiltonian with a
delta-function potential is pathological and should not be used for correct calculations, and we stress the point again here. The delta-function pseudo-potential was employed by Fermi, but it has been knwon for a long time that it is not consistent with scattering theory \cite{Esry}.

The Hamiltonian for a many-body Bose system with two-particle potential $U(|{\bf x}_j-{\bf x}_k|)$ is of the form:
\begin{equation}
H=-\sum_{j=1}^N \frac{\hbar^2}{2m}\Delta_j+\sum_{j<k}^N U(|{\bf x}_j-{\bf x}_k|)\ .
\label{1}
\end{equation}
The Hartree-Fock wavefunction describing Bose particles is given by the product of one-particle wavefunctions as
\begin{equation}
\Psi({\bf x}_1,{\bf x}_2,...,{\bf x}_N;t)=\prod_{j=1}^{N} \psi_N({\bf x}_j,t)\ .
\label{2}
\end{equation}
The standard variational procedure with this trial wavefunction used in the Hamiltonian (\ref{1}) yields the time-dependent Hartree-Fock equation for the one-particle wavefunction $\psi_N({\bf x},t)$ as
\begin{equation}
i\hbar\frac{\partial}{\partial t}\psi_N({\bf x},t)=\left\{-\frac{\hbar^2}{2m}\Delta+U_{HF}({\bf x},t)\right\}\psi_N({\bf x},t)\ ,
\label{3}
\end{equation}
where $U_{HF}({\bf x},t)$ is the Hartree-Fock time-dependent mean-field potential given by  
\begin{equation}
U_{HF}({\bf x},t)=(N-1)\int U(|{\bf x}-{\bf x}'|)|\psi_N({\bf x}',t)|^2d{\bf x}'\ ,
\label{4}
\end{equation}
with the normalization
$\int |\psi_N({\bf x},t)|^2d{\bf x}=1$. 

In the stationary case one may use the ansatz
$\psi_N({\bf x},t)=e^{-(i/\hbar){\cal E}_N t}\phi_N({\bf x})$ where ${\cal E}_N$ is the marginal ground state energy. 
The stationary Hartree-Fock equation for the time-independent wavefunction $\phi_N({\bf x})$ is then
\begin{multline}
{\cal E}_N\phi_N({\bf x})=-\frac{\hbar^2}{2m}\Delta\phi_N({\bf x})\\ 
+(N-1)\left(\int U(|{\bf x}-{\bf x}'|)|\phi_N({\bf x}')|^2d{\bf x}'\right)\phi_N({\bf x})\ .
\label{5}
\end{multline}
The eigenenergy in Eq.~(\ref{5}) can be written in the form
\begin{equation}
{\cal E}_N=\langle K\rangle_N+(N-1)\langle U\rangle_N\ ,
\label{6}
\end{equation}
where $\langle K\rangle_N$ and $(N-1)\langle U\rangle_N$ are the average kinetic and potential terms defined by the integrals
\begin{gather}
\langle K\rangle_N=\int \phi_N^{*}({\bf x})\left(-\frac{\hbar^2}{2m}\Delta \right)\phi_N({\bf x})d{\bf x}\ ,
\label{7}\\
\langle U\rangle_N=\int\int U(|{\bf x}-{\bf x}'|)|\phi_N({\bf x})|^2|\phi_N({\bf x}')|^2d{\bf x}d{\bf x}'\ .
\label{8}
\end{gather}
The expectation value $E_N=\langle H\rangle_N$ of the Hamiltonian ({\ref 1}) with the trial wavefunction ({\ref 2}) is 
 \begin{equation}
E_N=N\langle K\rangle_N+\frac{N(N-1)}{2}\langle U\rangle_N\ .
\label{9}
\end{equation}
The expectation value $E_N$ can also be written from Eqs.~({\ref 6}) and ({\ref 9}) as
\begin{equation}
E_N=N{\cal E}_N-\frac{N(N-1)}{2}\langle U\rangle_N\ .  
\label{10}
\end{equation}

In the case $N\gg 1$ we have $\langle K\rangle_N\simeq \langle K\rangle_{N-1}$ and $\langle U\rangle_N\simeq \langle U\rangle_{N-1}$; it then follows from Eq.~(\ref{9}) and Eq.~(\ref{6}) that the energy of the last particle is $E_N-E_{N-1}= {\cal E}_N$, as expected. 

We define an effective potential $\tilde{U}(r)=0$ at $r<a_0$ and $\tilde{U}(r)=U(r)$ otherwise, where $a_0$ is the s-scattering length.
For a dilute Bose gas (see Appendix A) we can make the substitution $U(r)\rightarrow \tilde{U}(r)$ in Eq.~({\ref 3}),  since the region $r<a_0$ in the potential is not accessible for low energy (s-wave) particle scattering. 
We also introduce the condensate wavefunction $\psi({\bf x},t)=\sqrt{N}\psi_N({\bf x},t)$ and make the approximation (valid for a dilute gas) that this is slowly varying over the range of the effective potential
\begin{equation}
\int \tilde{U}(|{\bf x}-{\bf x}'|)|\psi({\bf x}',t)|^2d{\bf x}'\simeq |\psi({\bf x},t)|^2\int \tilde{U}(|{\bf x}-{\bf x}'|)d{\bf x}'\ .
\label{11}
\end{equation}

In the thermodynamic limit, when the volume and the number of particles tend to infinity ($V\rightarrow \infty$,   $N\rightarrow \infty$) with fixed local density $|\psi({\bf x},t)|^2$,
the Hartree-Fock equation ({\ref 3}) with the approximation ({\ref 11}) yields the GP equation
\begin{equation}
i\hbar\frac{\partial}{\partial t}\psi({\bf x},t)=\left(-\frac{\hbar^2}{2m}\Delta+g_0|\psi({\bf x},t)|^2\right)\psi({\bf x},t)\ ,
\label{12}
\end{equation}
with the coupling parameter $g_0$ given by
\begin{equation}
g_0= \int \tilde{U}(|{\bf x}|)d{\bf x}= 4\pi \int_{a_0}^{\infty} U(r)r^2dr\ ,
\label{13}
\end{equation}
where we have used the effective potential cut-off as defined above. 
The normalization condition is $\int_V |\psi({\bf x},t)|^2d{\bf x}=N$.

 In the modified Born approximation (MBA)  (see Appendix A) the s-scattering length is given by $a_0=(m/\hbar^2)\int_{a_0}^{\infty} U(r)r^2dr$.  
 Combined with Eq.~(\ref{13}) this leads to the well-known formula  for the coupling parameter
\begin{equation}
g_0= \frac{4\pi a_0\hbar^2}{m}\ ,
\label{14}
\end{equation}
which is correct for a nearly ideal Bose gas \cite{Pit,Peth,Griff} with positive s-scattering length. 

We emphasize that this derivation of the GP equation with the coupling parameter in Eq.~(\ref{14}) does not use the standard Fermi pseudo-potential ansatz $\tilde{U}({\bf x})=(4\pi a_0\hbar^2/m)\delta({\bf x})$. 
Furthermore,  the MBA method applied here leads to a self-consistent definition of the s-scattering length $a_0$ that does not suffer from the divergence otherwise encountered for the Lennard-Jones or other potentials (see Appendix A).

Note that in the GP equation the s-scattering length $a_0$ should be positive (i.e.\ the effective potential must be repulsive), otherwise the sound velocity $c=\sqrt{g_0n/m}$ becomes imaginary. Moreover, both the GP equation (\ref{12}) and the Bogoliubov theory are applicable only under the two conditions that $\sqrt{a_0^3n}\ll1$ and $ka_0\ll1$ where $k$ is the wave number; the second condition is connected with the fact that the interaction of the particles is described only by the s-scattering waves.

\section{Bose-Einstein condensation in dense Bose fluids} 

In this section, using some approximations, we derive a full set of field equations describing the BEC in both a Bose gas and liquid $^4$He. Our approach is based on the Heisenberg equations for boson annihilation and  creation field operators in a Fock space. The Hamiltonian for the boson system including the interatomic potential can be written in terms of annihilation and  creation field operators $\hat{\psi}({\bf x},t)$ and $\hat{\psi}^{\dagger}({\bf x},t)$ in the form
\begin{multline}
\hat{H}=\int \hat{\psi}^{\dagger}({\bf x})\left(-\frac{\hbar^2}{2m}\Delta \right)\hat{\psi}({\bf x})d{\bf x}
\\ +\frac{1}{2}\int\int \hat{\psi}^{\dagger}({\bf x})\hat{\psi}^{\dagger}({\bf x}')U(|{\bf x}-{\bf x}'|)\hat{\psi}({\bf x}')\hat{\psi}({\bf x})d{\bf x}d{\bf x}'\ .
\label{31}
\end{multline}
This Hamiltonian yields the Heisenberg equation for the time-dependent field operator $\hat{\psi}({\bf x},t)$ as
\begin{multline}
i\hbar\frac{\partial}{\partial t}\hat{\psi}({\bf x},t)=-\frac{\hbar^2}{2m}\Delta\hat{\psi}({\bf x},t)
\\ +\left(\int U(|{\bf x}-{\bf x}'|)\hat{\psi}^{\dagger}({\bf x}',t)\hat{\psi}({\bf x}',t)d{\bf x}'\right)\hat{\psi}({\bf x},t)\ .
\label{32}
\end{multline}
Without loss of generality we may write the interatomic potential $U(r)$ as a sum of three potentials of differing ranges,
\begin{equation}
U(r)=\tilde{U}_s(r)+\tilde{U}_c(r)+\tilde{U}_l(r)\ .
\label{33}
\end{equation}
Here $\tilde{U}_s(r)$ is a short-range potential: $\tilde{U}_s(r)=U(r)$ for $r<a_s$ and $\tilde{U}_s(r)=0$ otherwise.
The parameter $a_s$ plays a similar role to that of $a_0$ in the nearly-ideal case treated in the previous section, being in effect the closest distance that the particles can actually approach one another.  So for a dilute gas $a_s\rightarrow a_0$, while for a denser, more strongly interacting fluid we will have $a_s<a_0$.
The potential $\tilde{U}_c(r)$ is medium-range, $\tilde{U}_c(r)=U(r)$ for $r\in (a_s,a_c)$  (with $a_c>a_s$) and $\tilde{U}_c(r)=0$ otherwise, while $\tilde{U}_l(r)$ is a long-range potential given by $\tilde{U}_l(r)=U(r)$ for $r>a_c$ and $\tilde{U}_l(r)=0$ otherwise.  The parameter $a_c$ at which these two are divided characterises the maximum range of significant correlations in the fluid; points further away than this are taken to have influence only via their bulk average values, but properties of the fluid at closer points may depend on local structure. For a dilute gas $a_c\rightarrow \infty$.
The values of both $a_s$ and $a_c$ depend on the density in the bulk.

By definition, the potential $\tilde{U}_s(r)$ describes the hard core region of the potential $U(r)$ that is forbidden to the particles at low temperatures, and hence we can neglect this part of the potential in Eq.~(\ref{32}). Thus we can write the expression in the parentheses in Eq.~(\ref{32}) as
\begin{multline}
\int U(|{\bf x}-{\bf x}'|)\hat{N}({\bf x}',t)d{\bf x}'\\
=\int \tilde{U}_c(|{\bf x}-{\bf x}'|)\hat{N}({\bf x}',t)d{\bf x}'
+{\cal U}({\bf x},t)+\\
\int \tilde{U}_{l}(|{\bf x}-{\bf x}'|)(\hat{N}({\bf x}',t)-\langle\hat{N}({\bf x}',t)\rangle)d{\bf x}'\ ,
\label{34}
\end{multline}
where $\hat{N}({\bf x},t)=\hat{\psi}^{\dagger}({\bf x},t)\hat{\psi}({\bf x},t)$ and the potential ${\cal U}({\bf x},t)$ is defined as
\begin{equation}
{\cal U}({\bf x},t)=\int \tilde{U}_{l}(|{\bf x}-{\bf x}'|)\langle\hat{\psi}^{\dagger}({\bf x}',t)\hat{\psi}({\bf x}',t)\rangle d{\bf x}'\ .
\label{35}
\end{equation}
Here $\langle ...\rangle={\rm Tr}(...\rho_0)$ where $\rho_0$ is the density operator at an initial time $t=0$.
Approximating the first term on the right side of Eq.~(\ref{34}) by the product ${[\int \tilde{U}_c(|{\bf x}-{\bf x}'|)d{\bf x}']\hat{\psi}^{\dagger}({\bf x},t)\hat{\psi}({\bf x},t)}$, and neglecting in Eq.~(\ref{34}) the last small term describing the fluctuation of the potential ${\cal U}$, we can write the Heisenberg equation (\ref{32}) for the field operator $\hat{\psi}({\bf x},t)$ as approximately
\begin{equation}
i\hbar\frac{\partial}{\partial t}\hat{\psi}=-\frac{\hbar^2}{2m}\Delta\hat{\psi}+{\cal U}\hat{\psi}+G\hat{\psi}^{\dagger}\hat{\psi}\hat{\psi}\ ,
\label{36}
\end{equation}
where ${\cal U}={\cal U}({\bf x},t)$ and the coupling parameter $G$ is 
\begin{equation}
G=\int \tilde{U}_c(|{\bf x}-{\bf x}'|)d{\bf x}'=4\pi\int_{a_s}^{a_c} U(r)r^2dr\ .
\label{37}
\end{equation}
 
We note that Eq.~(\ref{36}) is the Heisenberg equation for the effective Hamiltonian given by 
\begin{multline}
\hat {H}=\int \hat{\psi}^{\dagger}({\bf x})\left(-\frac{\hbar^2}{2m}\Delta +{\cal U}({\bf x})\right)\hat{\psi}({\bf x})d{\bf x}\\ 
+\frac{1}{2}\int G\hat{\psi}^{\dagger}({\bf x})\hat{\psi}^{\dagger}({\bf x})\hat{\psi}({\bf x})\hat{\psi}({\bf x})d{\bf x}\ .
\label{38}
\end{multline}
We decompose the field operator as $\hat{\psi}({\bf x},t)=\phi({\bf x},t)+\hat{\eta}({\bf x},t)$ where $\phi({\bf x},t)$ is the mean field $\langle \hat{\psi}({\bf x},t)\rangle$, and hence the average value of the field operator $\hat{\eta}({\bf x},t)$ vanishes: $\langle \hat{\eta}({\bf x},t)\rangle=0$. That is, $\hat{\eta}({\bf x},t)$ describes quantum and thermal fluctuations around the `condensate wave function' $\phi({\bf x},t)$.  

The last term in Eq.~(\ref{36}) is proportional to the product of three field operators and it can be written as
\begin{equation}
\hat{\psi}^{\dagger}\hat{\psi}\hat{\psi}=|\phi|^2\phi+2|\phi|^2\hat{\eta}+\phi^2\hat{\eta}^{\dagger}+2\phi\hat{\eta}^{\dagger}\hat{\eta}+\phi^{*}\hat{\eta}\hat{\eta}+\hat{\eta}^{\dagger}\hat{\eta}\hat{\eta}\ .
\label{39}
\end{equation}
We use below the approximate decomposition of the product of three time-dependent field operators
in the last term of Eq.~(\ref{39}) in the form
\begin{equation}
\hat{\eta}^{\dagger}\hat{\eta}\hat{\eta}=
\bcontraction{}{\hat{\eta}^{\dagger}}{}{\hat{\eta}} \hat{\eta}^{\dagger}\hat{\eta}\hat{\eta} 
+\bcontraction{}{\hat{\eta}^{\dagger}}{\hat{\eta}}{\hat{\eta}} \hat{\eta}^{\dagger}\hat{\eta}\hat{\eta}=2\langle\hat{\eta}^{\dagger}\hat{\eta}\rangle\hat{\eta}\ ,
\label{40}
\end{equation}
where the pairing is given by
$\bcontraction[0.7ex]{}{\hat{\eta}^{\dagger}}{}{\hat{\eta}} \hat{\eta}^{\dagger}\hat{\eta}
=\langle\hat{\eta}^{\dagger}\hat{\eta}\rangle$. 
Using this approximation, which implies also neglecting in Eq.~(\ref{39}) the penultimate term $\phi^{*}\hat{\eta}\hat{\eta}$
(because it is proportional to $\hat{\eta}^2$) we can rewrite Eq.~(\ref{36}) as
\begin{multline}
i\hbar\frac{\partial}{\partial t}\hat{\psi}=-\frac{\hbar^2}{2m}\Delta\hat{\psi}+{\cal U}\hat{\psi}+G|\phi|^2\phi\\
+2G(|\phi|^2+\langle\hat{\eta}^{\dagger}\hat{\eta}\rangle)\hat{\eta}+G\phi^2\hat{\eta}^{\dagger}+2G\phi\hat{\eta}^{\dagger}\hat{\eta}\ .
\label{41}
\end{multline}
The expectation of Eq.~(\ref{41}) yields the wave equation for the condensate wave function $\phi({\bf x},t)$
\begin{equation}
i\hbar\frac{\partial}{\partial t}\phi=-\frac{\hbar^2}{2m}\Delta\phi+{\cal U}\phi+2G\langle \hat{\eta}^{\dagger}\hat{\eta}\rangle\phi+G|\phi|^2\phi\ .
\label{42}
\end{equation}
Subtracting Eq.~(\ref{41}) from Eq.~(\ref{42}) then leads to the approximate Heisenberg equation for the operator $\hat{\eta}({\bf x},t)$
\begin{eqnarray}
i\hbar\frac{\partial}{\partial t}\hat{\eta}=-\frac{\hbar^2}{2m}\Delta\hat{\eta}+{\cal U}\hat{\eta}+2G(|\phi|^2+\langle \hat{\eta}^{\dagger}\hat{\eta}\rangle)\hat{\eta}
\nonumber \\ +G\phi^2\hat{\eta}^{\dagger} +2G\phi(\hat{\eta}^{\dagger}\hat{\eta}-\langle \hat{\eta}^{\dagger}\hat{\eta}\rangle)\ .
\label{43}
\end{eqnarray}
Neglecting the last small term proportional to the operator fluctuations $\hat{\eta}^{\dagger}\hat{\eta}-\langle \hat{\eta}^{\dagger}\hat{\eta}\rangle$ we find the linear equation for the operators $\hat{\eta}$ and $\hat{\eta}^{\dagger}$ in the form
\begin{equation}
i\hbar\frac{\partial}{\partial t}\hat{\eta}=-\frac{\hbar^2}{2m}\Delta\hat{\eta}+{\cal U}\hat{\eta}
+2G(|\phi|^2+\langle \hat{\eta}^{\dagger}\hat{\eta}\rangle)\hat{\eta}+G\phi^2\hat{\eta}^{\dagger}\ .
\label{44}
\end{equation}

The full system of equations describing the dense Bose gas and liquid $^4$He consists of Eq.~(\ref{42}) and its conjugate, and Eq.~(\ref{44}) and its adjoint. 

Using the decomposition  $\hat{\psi}({\bf x},t)=\phi({\bf x},t)+\hat{\eta}({\bf x},t)$, where the field $\phi$ describes the condensate and the operator field $\hat{\eta}$ describes the excitation of the condensate, one can write the full density $n=\langle\hat{\psi}^{\dagger}\hat{\psi}\rangle$ of the Bose system in the form 
\begin{equation}
n({\bf x},t)=n_c({\bf x},t)+n_{ex}({\bf x},t)\ ,
\label{45}
\end{equation}
where $n_c$ is the density of the condensate and $n_{ex}$ is the density of the `excited particles' given by
\begin{equation}
n_c({\bf x},t)=|\phi({\bf x},t)|^2,~~~n_{ex}({\bf x},t)=\langle \hat{\eta}^{\dagger}({\bf x},t)\hat{\eta}({\bf x},t)\rangle\ .
\label{46}
\end{equation}

For constant $n$ the potential ${\cal U}$ in Eq.~(\ref{35}) is
\begin{equation}
{\cal U}=G_ln,~~~G_l=4\pi\int_{a_c}^{\infty} U(r)r^2dr\ .
\label{47}
\end{equation} 
We also introduce another coupling parameter $g=4\pi\int_{a_s}^{\infty} U(r)r^2dr$ which can be written in the form
\begin{equation}
g=4\pi\int_{a_s}^{a_c} U(r)r^2dr+4\pi\int_{a_c}^{\infty} U(r)r^2dr=G+G_l\ .
\label{48}
\end{equation}

Eqs.~(\ref{47}) and (\ref{48}) lead to the potential ${\cal U}=(g-G)n$ where the coupling parameters $g=g(n)$ and $G=G(n)$ are functions of density $n$, since the parameters $a_s(n)$ and $a_c(n)$ depend on density.
Thus the system of wave equations given by Eq.~(\ref{42}) and (\ref{44}) for the condensate wave function $\phi$  and the field operator $\hat{\eta}$ describing the excitation of the condensate becomes
\begin{equation}
i\hbar\frac{\partial}{\partial t}\phi=-\frac{\hbar^2}{2m}\Delta\phi+{\cal V}\phi+G|\phi|^2\phi\ ,
\label{49}
\end{equation}
\begin{equation}
i\hbar\frac{\partial}{\partial t}\hat{\eta}=-\frac{\hbar^2}{2m}\Delta\hat{\eta}+\tilde{{\cal V}}\hat{\eta}
+G\phi^2\hat{\eta}^{\dagger}\ .
\label{50}
\end{equation}
The potentials in this coupled system of equations are
\begin{equation}
{\cal V}(n)=(g-G)n+2Gn_{ex},~\tilde{{\cal V}}(n)=(g+G)n\ ,
\label{51}
\end{equation}
where we emphasize again that the coupling parameters $G(n)$ and $g(n)$ are functions of the density $n$. 
In the general case, that of liquid $^4$He or a dense Bose gas, these two parameters are different.
However, for low enough densities, when $\sqrt{a_s^3n}\ll1$, we may take $a_c\rightarrow\infty$ and hence $G(n)=g(n)$ (though not necessarily $g(n)=g_0$).  We call this regime that of the \emph{dilute Bose gas} (DBG): it is described by the system of Eqs.~(\ref{49}) and (\ref{50}) with $G(n)=g(n)$, ${\cal V}(n)=2g(n)n_{ex}$ and $\tilde{{\cal V}}(n)=2g(n)n$.

For lower densities still, we have a \emph{nearly ideal Bose gas} (NIBG), in which $\sqrt{a_0^3n}\ll1$. Not only does $a_c\rightarrow\infty$, but now $a_s=a_0$, and hence $G=g=g_0$. In this case we have Eq.~(\ref{49}) with the parameters ${\cal V}=2g_0n_{ex}$ and $G=g_0$; this does go over to the GP equation (\ref{12}) in the limiting case when $n_{ex}\rightarrow 0$, but the coupled system of  Eqs.~(\ref{49}) and (\ref{50}) differs from the Bogoliubov theory (see Sec. VI and Appendix C, and the discussion in Sec. VIII). 
 
We note that Eq.~(\ref{49}) for the condensate wave function can also be written in the functional form
\begin{equation}
i\hbar\frac{\partial}{\partial t}\phi=\frac{\delta {\cal H}_c}{\delta\phi^{*}}\ .
\label{52}
\end{equation}
Here the energy ${\cal H}_c$ is the functional of the condensate wave function $\phi$ given by
\begin{equation}
{\cal H}_c=\int\left[-\frac{\hbar^2}{2m}(\phi^{*}\Delta\phi)+{\cal V}|\phi|^2+\frac{G}{2}|\phi|^4 \right]d{\bf x}\ .
\label{53}
\end{equation}

We may consider, as an example, propagation of superfluid $^4$He in a channel. Assuming that the channel is parallel to the $z$ axis, the solution of Eq.~(\ref{49}) is
\begin{equation}
\phi(z,t)=\sqrt{n_c}\exp[i(k_0z-\omega_0t)+i\theta]\ ,
\label{54}
\end{equation}
where the wave number $k_0$ and the frequency $\omega_0$ are 
\begin{equation}
\hbar k_0=mv,\quad\hbar\omega_0=\hbar\Omega+\frac{mv^2}{2}\ ,
\label{55}
\end{equation}
with $\hbar\Omega=gn+Gn_{ex}$;
the velocity is defined by ${\bf v}=(\hbar/m)\nabla \Theta$, where $\Theta$ is the phase of the wave function $\phi$. 
Thus Eq.~(\ref{54}) shows that the superfluid $^4$He propagates along the channel as a plane wave with the wave number $k_0$ and frequency $\omega_0$ defined in Eq.~(\ref{55}).

\section{Phonon spectrum and ground state energy in bose fluids}

The phonon spectrum in a Bose fluid can be derived 
from Eq.~(\ref{49}) without requiring a small density parameter $\sqrt{a_0^3n}$. The solution of this equation for a homogeneous condensate field $\phi(t)$ with both $n_c$ and $n$ constant is
\begin{equation}
\phi(t)=\sqrt{n_c}\exp[-i\Omega t+i\theta]\ ,~~~\hbar\Omega=gn+Gn_{ex}\ .
\label{56}
\end{equation}
Using this we define the ansatz $\phi=\tilde{\phi}\exp[-i\Omega t+i\theta]$ which transforms Eq.~(\ref{49}) to the wave equation 
\begin{equation}
i\hbar\frac{\partial}{\partial t}\tilde{\phi}=-\frac{\hbar^2}{2m}\Delta\tilde{\phi}-Gn_c\tilde{\phi}+G|\tilde{\phi}|^2\tilde{\phi}\ .
\label{57}
\end{equation}
This has the stationary solution $\tilde{\phi}=\sqrt{n_c}$, around which there are small fluctuations $F({\bf x},t)$,  
\begin{equation}
\tilde{\phi}({\bf x},t)=\sqrt{n_c}+F({\bf x},t)\ ,
\label{58}
\end{equation}
the small perturbations satisfy the linearized equation
\begin{equation}
i\hbar\frac{\partial}{\partial t}F=-\frac{\hbar^2}{2m}\Delta F+Gn_cF+Gn_cF^{*}\ .
\label{59}
\end{equation}
 
The general solution of the linear equation (\ref{59}) is
\begin{equation}
F({\bf x},t)=\frac{1}{\sqrt{V}}\sum_{\bf k}\left[u_{\bf k}e^{i({\bf k}{\bf x}-\omega_kt)}+v_{\bf k}^{*}e^{-i({\bf k}{\bf x}-\omega_kt)}\right]\ ,
\label{60}
\end{equation}
where $V=L^3$ is the quantization volume, ${\bf k}$ is the discrete wave number ${\bf k}=2\pi {\bf n}/L$, and
$\bf n$ is the vector with the components $0,\pm 1,\pm2,...$~.
The substitution of the decomposition given by Eq.~(\ref{60}) into Eq.~(\ref{59}) yields the system of equations
\begin{align}
(E_k-\varepsilon_k-Gn_c)u_{\bf k}&=Gn_cv_{\bf k}\ ,
\label{61}\\
(E_k+\varepsilon_k+Gn_c)v_{\bf k}&=-Gn_cu_{\bf k}\ ,
\label{62}
\end{align}
where $E_k$ and and $\varepsilon_k$ are respectively the excitation energy and the free particle energy, given by
\begin{equation}
E_k=\hbar\omega_k\ ,\quad\varepsilon_k=\frac{\hbar^2k^2}{2m}\ .
\label{63}
\end{equation}
The non-zero solution of Eqs.~(\ref{61}) and (\ref{62}) leads to the excitation energy being
\begin{equation}
E_k=\sqrt{2Gn_c\varepsilon_k+\varepsilon_k^2}\ .
\label{64}
\end{equation}

From Eq.~(\ref{64}) we may find the phonon velocity $c$ using the standard relation
\begin{equation}
c=\lim_{k\rightarrow 0}\frac{1}{\hbar} \frac{\partial E_k}{\partial k}=\sqrt{\frac{Gn_c}{m}}\ .
\label{65}
\end{equation}
Thus the energy spectrum of the elementary excitations and the coupling parameter $G$ can be written as
\begin{equation}
E_k=\sqrt{(c\hbar k)^2+\varepsilon_k^2}\ ,\quad G=\frac{mc^2}{n_c}\ .
\label{66}
\end{equation}
The energy spectrum of elementary excitations in Eqs.~(\ref{64}) and (\ref{66}) has the same form as the Bogoliubov spectrum for a dilute Bose gas, but has been found without assuming low density or weak coupling for the fluid. 

The ground state wavefunction $\phi_0$ of an arbitrary Bose fluid is the $c$-number solution to Eq.~(\ref{36}):
\begin{equation}
i\hbar\frac{\partial}{\partial t}\phi_0=-\frac{\hbar^2}{2m}\Delta\phi_0+{\cal U}\phi_0+G|\phi_0|^2\phi_0\ ,
\label{67}
\end{equation}
where ${\cal U}=(g-G)n$ is the potential and the normalization condition  is given by $|\phi_0|^2=n_c$; 
the density of excited particles $n_{ex}$ and hence the condensate density $n_c$ defining the normalization can be found from Eq.~(\ref{50}) (see also Sec. VI). Note that Eq.~(\ref{67}) is equal to Eq.~(\ref{49}) for the condensate wavefunction $\phi$ only in the limiting case when $n_{ex}=0$. 

The homogeneous solution of Eq.~(\ref{67}) is the ground state wavefunction 
\begin{equation}
\phi_0(t)=\sqrt{n_c}\exp[-i\Omega_0 t+i\theta_0]\ ,\quad\hbar\Omega_0=gn-Gn_{ex}\ .
\label{68}
\end{equation}
Eq.~(\ref{67}) can also be written in the functional form of Eq.~(\ref{52}), replacing $\phi$ by $\phi_0$ and ${\cal H}_c$ by the ground state energy functional
\begin{equation}
E_0=\int_V \left[-\frac{\hbar^2}{2m}(\phi_0^{*}\Delta\phi_0) + {\cal U}|\phi_0|^2+\frac{G}{2}|\phi_0|^4\right]d{\bf x}\ .
\label{69}
\end{equation}
Here ${\cal U}=(g-G)n$ and the coupling parameters $G$ and $g$ are functions of the density $n$.
Eqs.~(\ref{68}) and (\ref{69}) lead to the ground state energy
\begin{equation}
E_0=\left[(g(n)-G(n))n_c+\frac{G(n)n_c^2}{2n}\right]N\ ,
\label{70}
\end{equation}
where  $N$ is the number of the particles in the volume $V$. 
At zero temperature the ground state wave function $\phi_0$ minimizes the ground state energy $E_0$, or equivalently the free energy ${\cal F}$, when the ground state energy is negative (see Sec. V).

For a dilute Bose gas (i.e.\ $\sqrt{a_s^3n}\ll1$ and $a_c\rightarrow\infty$, but $a_s(n)< a_0$) we have $G(n)=g(n)=mc^2/n_c$ but $g(n)\neq g_0$.  
From Eq.~(\ref{70}) the ground state energy per particle ${\cal E}_0=\lim_{N\rightarrow \infty}E_0/N$ is then
\begin{equation}
{\cal E}_0(n)=\frac{g(n)n_c^2}{2n}=\frac{mc^2n_c}{2n}\ ,
\label{71}
\end{equation}
where $c=\sqrt{g(n)n_c/m}$.


For the case of a nearly-ideal Bose gas we also have $a_s=a_0$ and hence $G=g=g_0$.
In this approximation Eqs.~(\ref{66}) and (\ref{153}) give 
\begin{equation}
g_0=4\pi\int_{a_0}^{\infty} U(r)r^2dr=\frac{4\pi a_0\hbar^2}{m}=\frac{mc^2}{n_c}\ ,
\label{73}
\end{equation}
and hence the coupling parameter $g_0= 4\pi a_0\hbar^2/m$ and the phonon velocity $c=(\hbar/m)\sqrt{4\pi a_0n_c}$. The ground state energy per particle is then
\begin{equation}
{\cal E}_0(n)=\frac{2\pi a_0\hbar^2n_c^2}{mn}\ .
\label{74}
\end{equation}
Note that the GP approximation is the limiting case of the NIBG 
where the condensate density is equal to the full density $n_c=n$.

For the Lennard-Jones potential the parameter $g(n)$ is (from Eq.~(\ref{48}))
\begin{equation}
g(n)=4\pi\int_{a_s}^{\infty} U(r)r^2dr=Q\left[\left(\frac{r_0}{a_s}\right)^9-3\left(\frac{r_0}{a_s}\right)^3    \right]\ ,
\label{75}
\end{equation}
where $Q=16\pi\epsilon r_0^3/9$ and $a_s(n)$ is density dependent.
For a NIBG this result combined with the relation $g=g_0$ leads to Eq.~(\ref{155}) (see Appendix A). 

Similarly, from Eqs.~(\ref{37}) and (\ref{75}), the general result for the coupling parameter $G(n)$ for the Lennard-Jones potential is 
\begin{equation}
G(n)=g(n)+Q\left[3\left(\frac{r_0}{a_c}\right)^3-\left(\frac{r_0}{a_c}\right)^9\right]\ ,
\label{76}
\end{equation}
where $a_c(n)$ depends on the density.
 

\section{Condensation of liquid $^4$He at zero temperature}

We derive in this section the ground state energy and the condensation fraction of liquid $^4$He for a wide range of densities. These results are based on minimizing the free energy, which coincides with the ground state energy at zero temperature; the majority of equations in this section apply only to Bose fluids with negative ground state energy at zero temperature. 

In the case when $T\rightarrow 0$ the partition function is just ${\cal Z}=e^{-E_0/\Theta}$ (where $\Theta=k_BT$) and the free energy and the entropy are ${\cal F}=-\Theta{\rm ln}{\cal Z}$ and $S=-k_B{\cal P}_0{\rm ln}{\cal P}_0$ where ${\cal P}_0={\cal Z}^{-1}e^{-E_0/\Theta}$ is the probability of the ground state. Hence in this limiting case ($T\rightarrow 0$) the free energy is equal to the ground state energy, ${\cal F}=E_0$, and the entropy is zero, $S=0$.

In the case of an interacting Bose fluid with negative ground state energy at $T=0$, 
the wavefunction $\phi_0$ of the ground state will minimize the free energy (i.e.\ the ground state energy given by Eqs.~(\ref{69}) and (\ref{70})) as a function of the condensate density $n_c$. 
In the thermodynamic limit (i.e.\ for a system of infinite extent at fixed density) the minimal principle can be expressed in terms of the energy per particle:
\begin{equation}
{\cal E}_0(n)=\min_{n_c}{\cal E}_0(n,n_c)\ ,
\label{77}
\end{equation}
which can be applied to liquid $^4$He at zero temperature. From Eq.~(\ref{70}) we have
\begin{equation}
{\cal E}_0(n,n_c)=-(G(n)-g(n))n_c+\frac{G(n)n_c^2}{2n}\ .
\label{78}
\end{equation}
Minimizing this with respect to the condensate density $n_c$ by setting $\partial {\cal E}_0(n,n_c)/\partial n_c =0$ yields the densities 
\begin{equation}
n_c=\left(1-\frac{g(n)}{G(n)}\right)n\ ,\quad n_{ex}=\frac{g(n)n}{G(n)}\ .
\label{79}
\end{equation}
Substituting back into Eq.~(\ref{78}) gives the ground state energy per particle
\begin{equation}
{\cal E}_0(n)=-\frac{n}{2G(n)}(G(n)-g(n))^2=-\frac{G(n)n_c^2}{2n}\ .
\label{80}
\end{equation}

The chemical potential of the condensate at $T=0$ can be calculated by definition as 
\begin{equation}
\mu=\frac{\partial E_0(N)}{\partial N}={\cal E}_0(n)+n\frac{\partial {\cal E}_0(n)}{\partial n}\ ,
\label{81}
\end{equation}
where $E_0(N)=N{\cal E}_0$.
The pressure at $T=0$ is $P=-\partial E_0(N)/\partial V$ which yields
\begin{equation}
P=-\frac{\partial {\cal E}_0(n)}{\partial v}=n^2\frac{\partial {\cal E}_0(n)}{\partial n}\ ,
\label{82}
\end{equation}
where $v=n^{-1}$. 
Combining Eqs.~(\ref{81}) and (\ref{82}) we find that the pressure at $T=0$ is given by equation $P=(\mu-{\cal E}_0)n$. Thus the chemical potential is 
\begin{equation}
\mu={\cal E}_0(n)+\frac{P(n)}{n}\ ,
\label{83}
\end{equation}
where we consider the pressure $P(n)$ to be a known function of the density $n$.

We note that Eqs.~(\ref{66}) and (\ref{79}) lead to
\begin{equation}
G(n)=g(n)+\frac{mc^2}{n}\ ,
\label{84}
\end{equation}
where $c$ is the (density-dependent) phonon velocity at $T=0$, giving $c=\sqrt{(G-g)n/m}$. 
Eqs.~(\ref{76}) and (\ref{84}) also yield the relation
\begin{equation}
mc^2=nQ\left[3\left(\frac{r_0}{a_c}\right)^3-\left(\frac{r_0}{a_c}\right)^9\right]\ ,\quad Q=\frac{16\pi\epsilon r_0^3}{9}\ ,
\label{85}
\end{equation}
where the parameter $a_c(n)$ depends on the density.

Combining Eqs.~(\ref{79}), (\ref{80}) and (\ref{84}) one can write the condensate and the excitation densities and the ground state energy  ${\cal E}_0(n)$ at $T=0$ in the form 
\begin{gather}
n_c=\frac{mc^2n}{mc^2+gn}\ ,\quad n_{ex}=\frac{gn^2}{mc^2+gn}\ ,
\label{86}\\
{\cal E}_0(n)=-\frac{m^2c^4}{2(mc^2+gn)}\ .
\label{87}
\end{gather}
Eqs.~(\ref{86}) and (\ref{87}) lead to the further relations
\begin{gather}
g(n)=\frac{mc^2n_{ex}}{nn_c}\ ,\quad G(n)=\frac{mc^2}{n_c}\ ,
\label{88}\\
{\cal E}_0(n)=-\frac{m^2c^4}{2nG}=-\frac{mc^2n_c}{2n}\ .
\label{89}
\end{gather}
Eqs.~(\ref{87}) and (\ref{89}) demonstrate that the ground state energy ${\cal E}_0(n)$ is negative, as expected in consequence of the minimal principle formulated in Eq.~(\ref{77}) for a zero-temperature Bose fluid of infinite extent.
They also show that the frequency introduced in Eq.~(\ref{56}) has the value
\begin{equation}
\hbar\Omega=2gn=\frac{2mc^2n_{ex}}{n_c}\ ,
\label{90}
\end{equation}
while the ground-state frequency $\Omega_0$ introduced in Eq.~(\ref{68}) is zero in consequence of 
Eq.~(\ref{79}).

To proceed further, we express the pressure in the bulk as a power series of the mass density $\rho=mn$:  
\begin{equation}
P=\beta_1\rho^2+\beta_2\rho^3+\beta_3\rho^4\ ,
\label{91}
\end{equation}
where $\beta_k=\alpha_k/m$ ($k=1,2,3$) and the resulting parameters in ${\rm c.g.s.}$ units are:
$\alpha_1/k_B=-54.08$, $\alpha_2/k_B=-545.7$, and $\alpha_3/k_B=6325$.
The parameters of this equation \cite{Dup,Dal} are found using Monte Carlo simulations \cite{Bor} and experimental data \cite{Abr,Bru}; they accurately represent the simulations and experimental data for the pressure, the sound velocity and the ground state energy over a wide range of densities (see also Appendix B).
The ground state energy can then be found by integration of Eq.~(\ref{82}), which yields  
\begin{equation}
{\cal E}_0=m\int (P/\rho^2)d\rho=\alpha_1\rho+\frac{1}{2}\alpha_2\rho^2+\frac{1}{3}\alpha_3\rho^3\ .
\label{92}
\end{equation}
Fig.~1 and Fig.~2 show the pressure and the ground state energy per particle given by Eqs.~(\ref{91}) and (\ref{92}), and also the same quantities according to an alternative but similar model given by Eqs.~(\ref{251}) and (\ref{252}).

\begin{figure}
\centering{}
  \includegraphics[width=9cm]{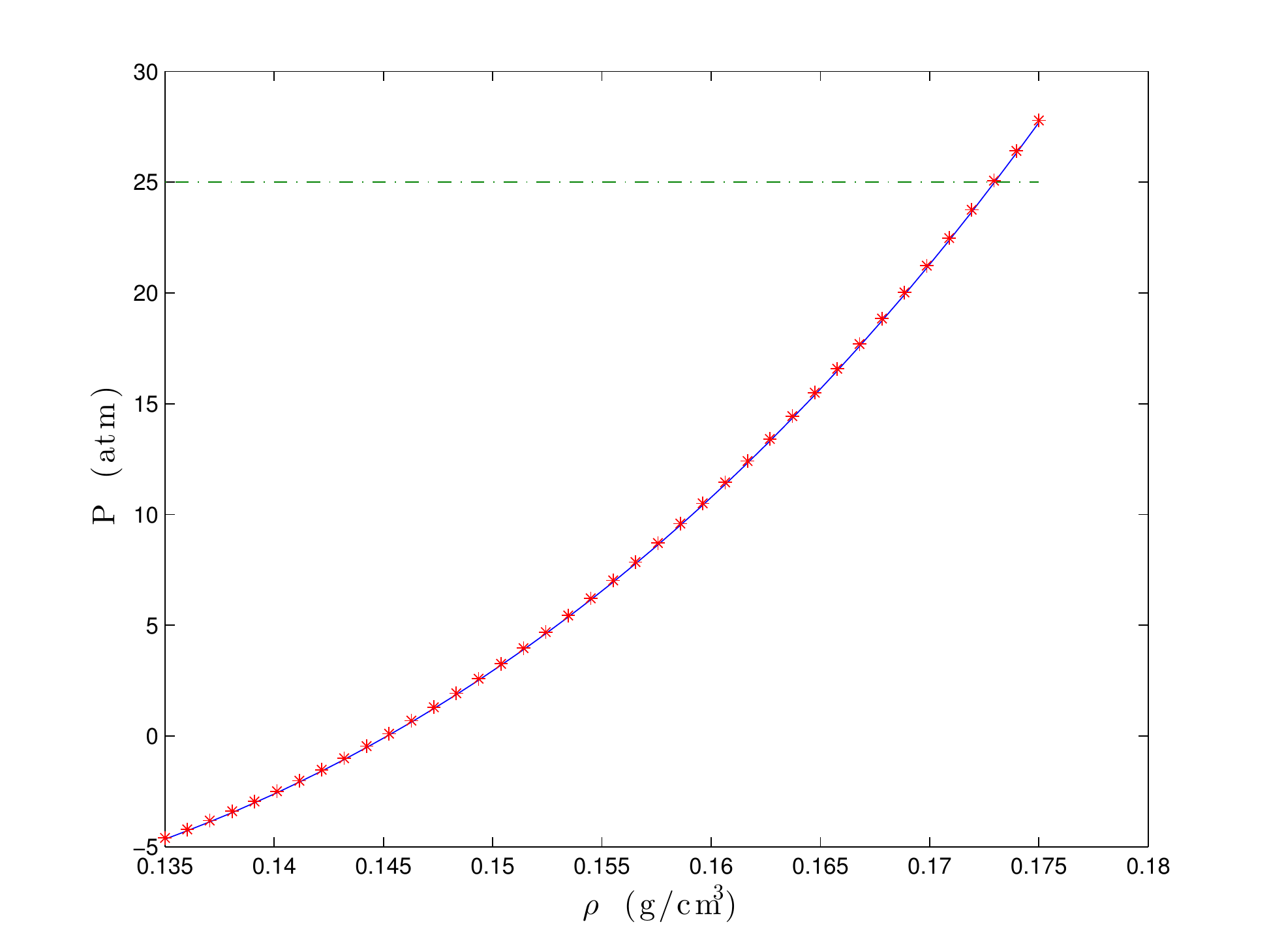}
  \caption{(Color online) Pressure as given by Eq.~(\ref{91}) (solid line) and Eq.~(\ref{251}) (stars). The dashed line indicates the melting pressure $P_m=25~ {\rm atm}$. }
  \label{F1}
\end{figure}
\begin{figure}
\centering{}
  \includegraphics[width=9cm]{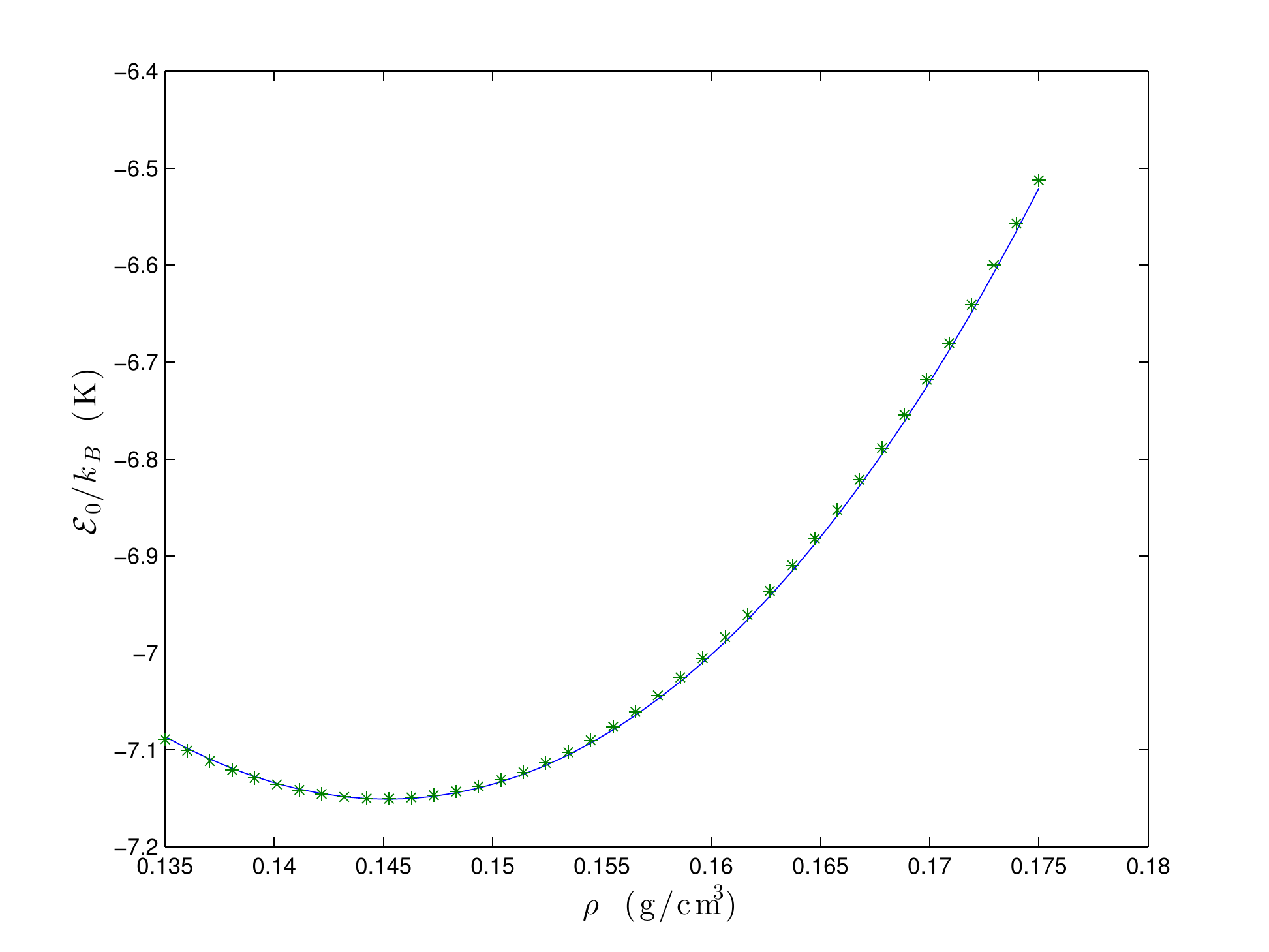}
  \caption{(Color online) Ground state energy per particle as given by Eq.~(\ref{92}) (solid line) and Eq.~(\ref{252}) (stars).}
  \label{F2}
\end{figure}

The sound velocity, $c=\sqrt{\partial P/\partial \rho}$, and the chemical potential in Eq.~(\ref{81}) are given by
\begin{gather}
c=(2\beta_1\rho+3\beta_2\rho^2+4\beta_3\rho^3)^{1/2},
\label{93}
\\
\mu=2\alpha_1\rho+\frac{3}{2}\alpha_2\rho^2+\frac{4}{3}\alpha_3\rho^3\ .
\label{94}
\end{gather}
In a zero temperature fluid, the sound velocity is in fact a true linear function of density, which yields another, more natural representation of the pressure, ground state energy, chemical potential and sound velocity (see Appendix B); nonetheless both representations are highly accurate. 
Fig.~3 and Fig.~4 present the chemical potential and sound velocity respectively according to both Eqs.~(\ref{94}) and (\ref{93}) and also Eqs.~(\ref{253}) and (\ref{250}).
\begin{figure}
\centering{}
  \includegraphics[width=9cm]{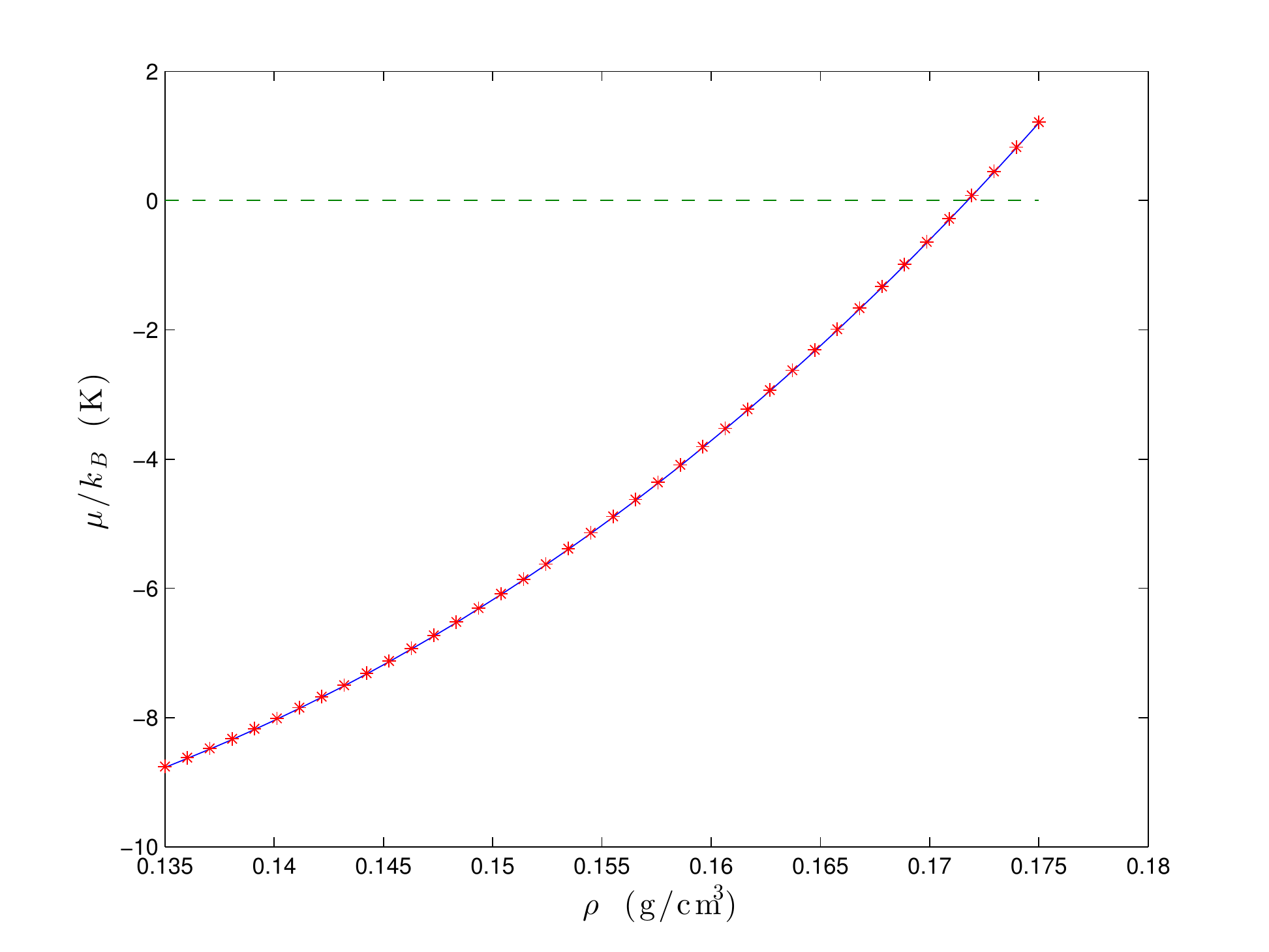}
  \caption{(Color online) Chemical potential given by Eq.~(\ref{94}) (solid line) and Eq.~(\ref{253}) (stars). The dashed line indicates the chemical potential $\mu=0$ at the melting pressure $P_m$ with $\rho =0.172~{\rm g~cm}^{-3}$.}
  \label{F3}
\end{figure}
\begin{figure}
\centering{}
  \includegraphics[width=9cm]{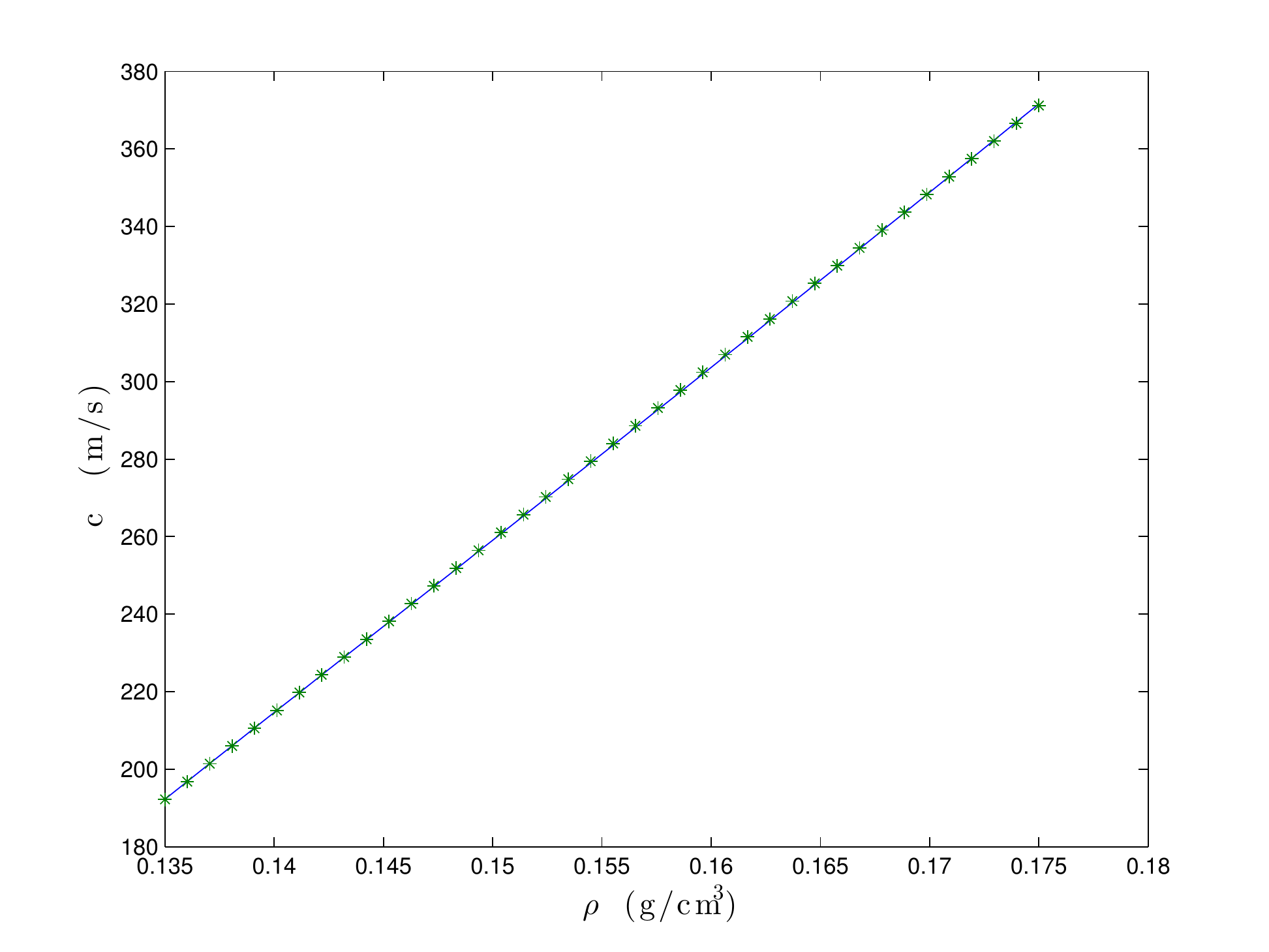}
  \caption{(Color online) Sound velocity given by Eq.~(\ref{93}) (solid line) and Eq.~(\ref{250}) (stars). }
  \label{F4}
\end{figure}

 Combining Eqs.~(\ref{88}), (\ref{89}), (\ref{92}) and (\ref{93}) we can present the coupling parameters $G$ and $g$ as functions of density $\rho$ in the form
\begin{equation}
\frac{G}{m}=-\frac{(2\alpha_1+3\alpha_2\rho+4\alpha_3\rho^2)^2}{2\alpha_1+\alpha_2\rho+(2/3)\alpha_3\rho^2}\ ,
\label{95}
\end{equation}
\begin{eqnarray}
\frac{g}{m}=  -\frac{(2\alpha_1+3\alpha_2\rho+4\alpha_3\rho^2)^2}{2\alpha_1+\alpha_2\rho+(2/3)\alpha_3\rho^2}   \nonumber \\   -2\alpha_1-3\alpha_2\rho-4\alpha_3\rho^2\ .
\label{96}
\end{eqnarray}
For example, these equations lead at $\rho=0.145~{\rm g~cm}^{-3}$ to the coupling parameters $G/(mk_B)=352~{\rm K~ cm^3~ g^{-1}}$ and $g/(mk_B)=166~{\rm K~ cm^3~ g^{-1}}$.

 The condensate fraction can be calculated by Eq.~(\ref{89}) as $\rho_c/\rho= -2{\cal E}_0/(mc^2)$. Thus using
 Eqs.~(\ref{92}) and (\ref{93}) we have the equation for the condensate fraction
\begin{equation}
\frac{\rho_c}{\rho}=-\frac{2\alpha_1+\alpha_2\rho+(2/3)\alpha_3\rho^2}{2\alpha_1+3\alpha_2\rho+4\alpha_3\rho^2}\ .
\label{97}
\end{equation}
The excitation fraction is then $\rho_{ex}/\rho=1-\rho_{c}/\rho$. For example, when $\rho=0.145~{\rm g~cm}^{-3}$ the condensate and excitation fractions are $\rho_c/\rho=0.528$ and $\rho_{ex}/\rho=0.472$ respectively.
The coupling parameters given by Eqs.~(\ref{95}), (\ref{96}) and the condensate and excitation fractions described by Eq.~(\ref{97}) are shown in Fig. 5 and Fig.6 respectively. We also derive in the Appendix B an alternative representation (with similar accuracy) for the coupling parameters $G$ and $g$ and for the fraction $\rho_c/\rho$.

\begin{figure}
\centering{}
  \includegraphics[width=9cm]{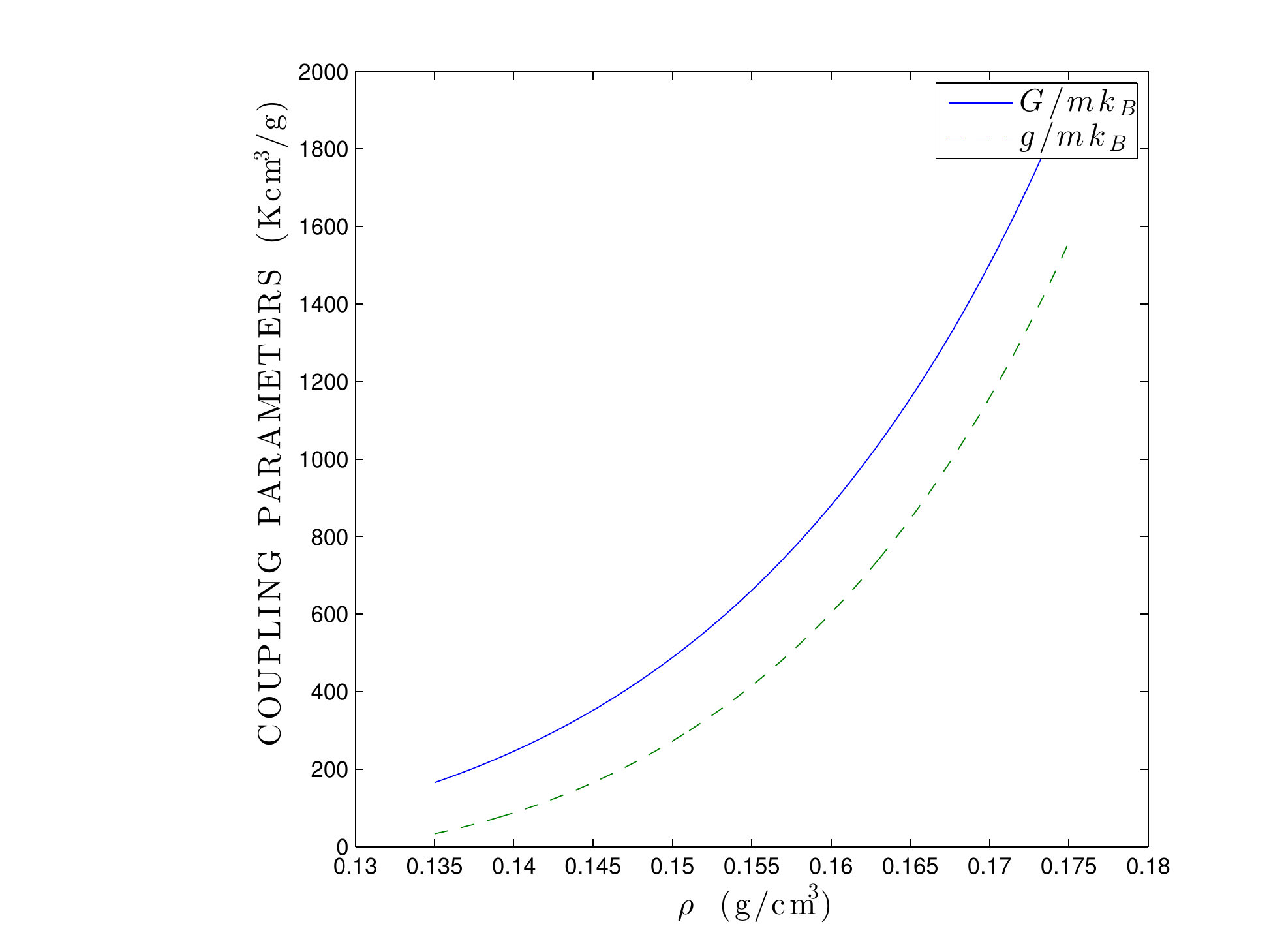}
  \caption{(Color online) Scaled coupling parameters $G$ (solid line) and $g$ (dashed line), given by Eq.~(\ref{95}) and Eq.~(\ref{96}). }
  \label{F5}
\end{figure}
\begin{figure}
\centering{}
  \includegraphics[width=9cm]{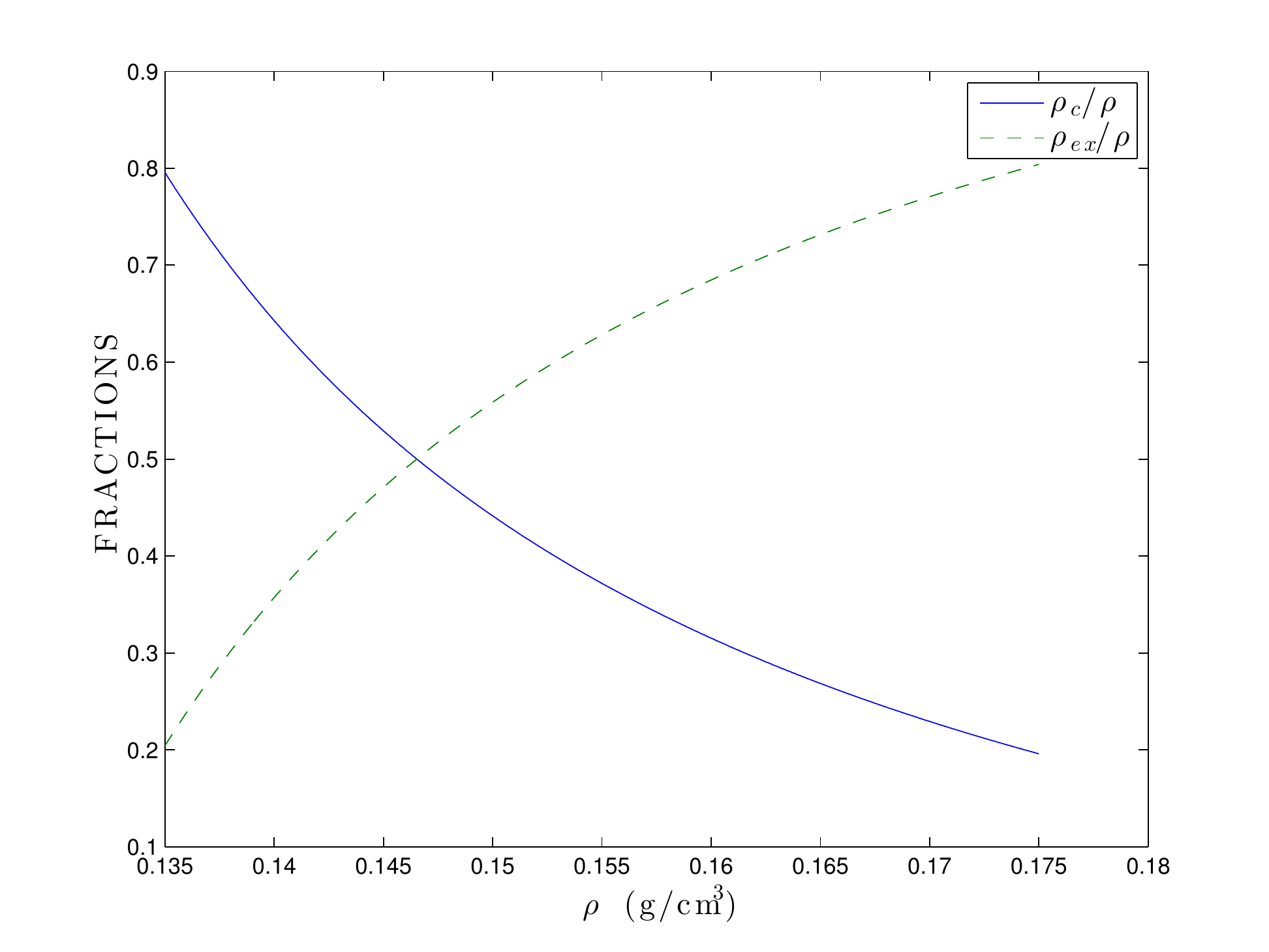}
  \caption{(Color online) Condensate and excitation fractions $\rho_c/\rho$ (solid line) and $\rho_{ex}/\rho$ (dashed line), given by Eq.~(\ref{97}).} 
  \label{F6}
\end{figure}

 \section{Phonon excitations and ground state energy in liquid $^4$He}

In this section we derive an effective Hamiltonian describing phonon excitations in liquid $^4$He.
Diagonalization of this effective Hamiltonian allows us to calculate the excitation and condensed densities and 
the ground state energy. Our approach also uses a cut-off procedure in momentum (or wavenumber) space; the cut-off parameter is determined by the second sound, and leads to convergent integrals.
 
The excitations of the BEC in liquid $^4$He can be described by Eq.~(\ref{50}).
 The last term in this equation is proportional to $\phi^2(t)$ and hence is explicitly time dependent term. We may exclude the explicit time dependence by introducing the new field operator 
\begin{equation}
\hat{\chi}({\bf x},t)=\hat{\eta}({\bf x},t)\exp(i\Omega t-i\theta)\ .
\label{98}
\end{equation}
Eq.~(\ref{50}) then yields the Heisenberg equation for the field operator $\hat{\chi}({\bf x},t)$
\begin{equation}
i\hbar\frac{\partial}{\partial t}\hat{\chi}=-\frac{\hbar^2}{2m}\Delta\hat{\chi}+Gn_c\hat{\chi}
+Gn_c\hat{\chi}^{\dagger}\ ,
\label{99}
\end{equation}
where the commutators for the fields $\hat{\chi}({\bf x},t)$ and $\hat{\chi}^{\dagger}({\bf x},t)$ are the same as the Bose fields $\hat{\eta}({\bf x},t)$ and $\hat{\eta}^{\dagger}({\bf x},t)$,
\begin{align}
[\hat{\chi}({\bf x},t),\hat{\chi}^{\dagger}({\bf x}',t)]&=\delta({\bf x}-{\bf x}')\ ,\nonumber\\
[\hat{\chi}({\bf x},t),\hat{\chi}({\bf x}',t)]&=0\ .
\label{100}
\end{align}

Eq.~(\ref{99}) can be written in the standard Heisenberg form as
\begin{equation}
i\hbar\frac{\partial}{\partial t}\hat{\chi}({\bf x},t)=[\hat{\chi}({\bf x},t),\hat{\cal H}]\ ,
\label{101}
\end{equation}
where the effective Hamiltonian is 
\begin{multline}
\hat{\cal H}=E+\int \hat{\chi}^{\dagger}({\bf x})\left(-\frac{\hbar^2}{2m}\Delta+Gn_c\right)\hat{\chi}({\bf x})d{\bf x} \\ +\frac{1}{2}\int Gn_c[\hat{\chi}({\bf x})\hat{\chi}({\bf x})
+\hat{\chi}^{\dagger}({\bf x})\hat{\chi}^{\dagger}({\bf x})]d{\bf x}\ ,
\label{102}
\end{multline}
$E$ is a constant connected with the ground state energy. 
The Hamiltonian (\ref{102}) can also be used to find the Heisenberg equation for the field operator $\hat{\eta}({\bf x},t)$. Since
\begin{equation}
\hat{\eta}({\bf x},t)=e^{i\hat{\cal H}t/\hbar}\hat{\chi}({\bf x},0)e^{-i\hat{\cal H}t/\hbar}e^{-i\Omega t+i\theta}\ ,
\label{103}
\end{equation}
it follows  by differentiation that
\begin{equation}
i\hbar\frac{\partial}{\partial t}\hat{\eta}({\bf x},t)=[\hat{\eta}({\bf x},t),\hat{\cal H}]+\hbar\Omega\hat{\eta}({\bf x},t)\ .
\label{104}
\end{equation}
The last term in Eq.~(\ref{104}) comes from the exponential factor in $\hat{\eta}({\bf x},t)=\hat{\chi}({\bf x},t)e^{-i\Omega t+i\theta}$. 

The Bose field $\hat{\chi}({\bf x})$ can quite generally be expressed as
\begin{equation}
\hat{\chi}({\bf x})=\frac{1}{\sqrt{V}}\sum_{\bf k}\left(\alpha_{\bf k}\hat{c}_{{\bf k}}+\beta_{\bf k}\hat{c}_{-{\bf k}}^{\dagger}\right)e^{i{\bf k}{\bf x}},
\label{105}
\end{equation}
where $V=L^3$ is the quantization volume and ${\bf k}$ is the discrete wave number ${\bf k}=2\pi {\bf n}/L$ (where
$\bf n$ is the vector with the components $0$, $\pm 1$, $\pm2$, \dots). Here $\hat{c}_{{\bf k}}$ and $\hat{c}_{{\bf k}}^{\dagger}$ are Bose annihilation and creation operators with the usual commutators $[\hat{c}_{{\bf k}},\hat{c}_{{\bf k}'}^{\dagger}]=\delta_{{\bf k},{\bf k}'}$ and $[\hat{c}_{{\bf k}},\hat{c}_{{\bf k}'}]=0$. 
Consistency with Eq.~(\ref{100}) requires
\begin{equation}
|\alpha_{\bf k}|^2-|\beta_{\bf k}|^2=1\ ,
\label{106}
\end{equation}
and since any complex phase can always be absorbed into the mode operators, we may take $\alpha_{\bf k}$ and $\beta_{\bf k}$ to be any real functions of the wavenumbers that satisfy Eq.~(\ref{106}). 

The effective Hamiltonian (\ref{102}) can be diagonalized in terms of  the mode operators introduced in Eq.~(\ref{105}) by an appropriate choice of $\alpha_{\bf k}$ and $\beta_{\bf k}$. The resulting diagonal form of the Hamiltonian is 
\begin{equation}
\hat{\cal H}=E_0+\sum_{\bf k}E_k\hat{c}_{{\bf k}}^{\dagger}\hat{c}_{{\bf k}}\ ,
\label{112}
\end{equation}
where the ground state energy $E_0$ and excitation energies $E_k$ can be written
\begin{align}
E_0=E+\sum_{\bf k}[(\varepsilon_k+Gn_c)\beta_{\bf k}^2+Gn_c\alpha_{\bf k}\beta_{\bf k}]\ ,
\label{108}\\
E_k = (\varepsilon_k+Gn_c)(\alpha_{\bf k}^2+\beta_{\bf k}^2)+2Gn_c\alpha_{\bf k}\beta_{\bf k}\ .
\label{107}
\end{align}
Here $\varepsilon_k={\hbar^2k^2}/{2m}$ is the energy of a free particle with momentum $p=\hbar k$.
The diagonalization condition leading to Eq.~(\ref{107}) is
\begin{equation}
(\varepsilon_k+Gn_c)\alpha_{\bf k}\beta_{\bf k}+\frac{1}{2}Gn_c(\alpha_{\bf k}^2+\beta_{\bf k}^2)=0\ .
\label{109}
\end{equation}
The solution of this with the condition Eq.~(\ref{106}) yields solutions for $\alpha_{\bf k}$ and $\beta_{\bf k}$ that can be written in the form
\begin{equation}
\alpha_{\bf k}=(1-\gamma_{\bf k}^2)^{-1/2}\ ,\quad\beta_{\bf k}=\gamma_{\bf k}(1-\gamma_{\bf k}^2)^{-1/2}\ .
\label{110}
\end{equation}
where the real function $\gamma_{\bf k}$ is
\begin{equation}
\gamma_{\bf k}=\frac{1}{mc^2}(E_k-\varepsilon_k-mc^2)\ ,\quad E_k=\sqrt{(c\hbar k)^2+ \varepsilon_k^2}\ ,
\label{111}
\end{equation}
where the sound velocity is defined as $c=\sqrt{Gn_c/m}$. 
Equivalently, we can satisfy Eq.~(\ref{106}) automatically by parameterising $\alpha_{\bf k} = \cosh\theta_{\bf k}$ and $\beta_{\bf k} = \sinh\theta_{\bf k}$ (so that $\gamma_{\bf k} = \tanh\theta_{\bf k}$) 
where $\exp(-4\theta_{\bf k})=1+2Gn_c/\varepsilon_k$.  Note that both $\theta_{\bf k}$ and $\gamma_{\bf k}$ are negative.


The zero-point energy $E$ in Eqs.~(\ref{102}) and (\ref{108}) is connected with the ground state energy $E_0$ by       
\begin{equation}
E_0=E+\delta E\ ,\quad\delta E=\frac{1}{2}\sum_{\bf k}(E_k-\varepsilon_k-mc^2)\ ,
\label{113}
\end{equation}
following from Eqs.~(\ref{108}), (\ref{110}) and (\ref{111}). 
For liquid $^4$He we also still have that the total ground state energy is given by $E_0=N{\cal E}_0(n)$ where  ${\cal E}_0(n)$ is defined in Eq.~(\ref{89}).

The energy spectrum $E_k=\sqrt{2Gn_c\varepsilon_k+\varepsilon_k^2}$ found by this diagonalization procedure is consistent with the previous result for the sound velocity and the spectrum given by Eq.~(\ref{65})
and Eq.~(\ref{66}), as shown in Eq.~(\ref{111}).
For sufficiently small wavenumber $k\ll mc/\hbar$
the energy of the excitations is a linear function of the wavenumbers, $E_k=c\hbar k$; 
in this regime the operators $\hat{c}_{{\bf k}}$ and $\hat{c}_{{\bf k}}^{\dagger}$ are phonon annihilation and creation operators.

Mathematically, our diagonalization procedure is equivalent to the well-known Bogoliubov canonical transformation \cite{Bog}; 
however, its physical basis differs from Bogoliubov's theory.  
Firstly, the field $\hat{\chi}({\bf x})$ is directly introduced as the fluctuating part of the interacting field, not by a canonical transformation from the free particle field. 
Secondly, our results are not explicitly perturbative; there are in the general case no small parameters in this theory. It is still true though that the nonhomogeneous part of the field must be in some sense `small'; in the presence of significant medium-scale structure (such as the roton clusters discussed in Sect.\ VII below) the explicit results are only accurate for small $k$ (more precisely, for $\hbar k\ll mc$), where the excitations are still phonon-like.

Using these results we can also write the time-dependent field operator $\hat{\chi}({\bf x},t)=e^{i\hat{\cal H}t/\hbar}\hat{\chi}({\bf x},0)e^{-i\hat{\cal H}t/\hbar}$ in an explicit form as
\begin{equation}
\hat{\chi}({\bf x},t)=\frac{1}{\sqrt{V}}\sum_{\bf k}\left(\frac{\hat{c}_{{\bf k}}e^{-iE_kt/\hbar}}{\sqrt{1-\gamma_{\bf k}^2}}+\frac{\hat{c}_{-{\bf k}}^{\dagger}\gamma_{\bf k}e^{iE_kt/\hbar}}{\sqrt{1-\gamma_{\bf k}^2}}\right)e^{i{\bf k}{\bf x}}\ .
\label{114}
\end{equation}
Eqs.~(\ref{98}), (\ref{114}) then lead to the correlation function
\begin{equation}
\langle\hat{\eta}^{\dagger}({\bf x}_2,t)\hat{\eta}({\bf x}_1,t)\rangle=\frac{1}{(2\pi)^{3}}\int N_{\bf k}e^{i{\bf k}({\bf x}_1-{\bf x}_2)}d{\bf k}\ .
\label{115}
\end{equation}
Here $N_{\bf k}$ is the distribution of the excited particles in momentum (or wavenumber) space given by
\begin{equation}
N_{\bf k}=\frac{n_{{\bf k}}+\gamma_{\bf k}^2(n_{{\bf k}}+1)}{1-\gamma_{\bf k}^2}\ ,\quad n_{\bf k}=\frac{1}{e^{\beta E_k}-1}\ ,
\label{116}
\end{equation}
where $n_{\bf k}=\langle\hat{c}_{{\bf k}}^{\dagger}   \hat{c}_{{\bf k}}\rangle$ is the phonon distribution. 
The anomalous correlation function follows from Eq.~(\ref{114}) as 
\begin{equation}
\langle\hat{\chi}({\bf x}_2,t)\hat{\chi}({\bf x}_1,t)\rangle=\frac{1}{(2\pi)^3}\int \frac{\gamma_{\bf k}(2n_{{\bf k}}+1)}{1-\gamma_{\bf k}^2}e^{i{\bf k}({\bf x}_2-{\bf x}_1)}d{\bf k}\ .
\label{117}
\end{equation}

From Eqs.~(\ref{115}) and (\ref{116}) the excitation density, ${n_{ex}=(2\pi)^{-3}\int N_{\bf k}d{\bf k}}$, can be written
\begin{equation}
n-n_c=\frac{1}{(2\pi)^{3}}\int \frac{
\gamma_{\bf k}^2}{1-\gamma_{\bf k}^2}d{\bf k}+\frac{1}{(2\pi)^{3}}\int \left(\frac{
1+\gamma_{\bf k}^2}{1-\gamma_{\bf k}^2}\right)n_{\bf k}d{\bf k}\ .
\label{118}
\end{equation}
The upper bound of the integrals in Eq.~(\ref{118}) needs careful consideration.  Although they do formally converge for large $k$, we still need to introduce a cutoff in momentum space on physical grounds.  The maximum value of momentum for which it is sensible to treat the excitations in liquid $^4$He as phonon-like is that corresponding to waves moving at the speed of the second sound, which at zero temperature is $c_0=c/\sqrt{3}$. 
This gives a cutoff at $k_c=mc_0/\hbar$. We note that the velocity $c_0$ is the maximal velocity of the waves in the subsystem of excitations \cite{Khal}.


Even in the region below the cutoff, small corrections can be made to the phonon spectrum $E_k$ given by Eq.~(\ref{111}). In the region $0<k<k_c$ a more accurate spectrum is
\begin{equation}
E_k=mc^2[q+\nu_3(\rho)q^3+\nu_5(\rho)q^5+...]\ ,\quad q=\frac{\hbar k}{mc}\ .                     
\label{119}
\end{equation}
Here the first term describes the linear spectrum $E_k=c\hbar k$ and the dimensionless coefficients $\nu_s(\rho)$ (with $s=3,5,...$) are small in the region $0<k<k_c$. We use in our calculations below only the first small correction to the energy spectrum given by Eq.~(\ref{119}) with $\nu_3\neq 0$, neglecting the other correction terms.
At $T=0$ with $\rho=0.145$ ${\rm g~cm}^{-3}$ the dimensionless parameter $\nu_3$ for liquid $^4$He can be estimated  from experimental data as $\nu_3\simeq -0.1$; alternatively, the same result can be obtained by a perturbative calculation that we shall present elsewhere. 

Substituting the expansion Eq.~(\ref{119}) into Eq.~(\ref{111}) gives
\begin{equation}
\gamma_{\bf k}=\gamma(q,\rho)=-1+q-\frac{1}{2}q^2+\nu_3(\rho)q^3+\nu_5(\rho)q^5+...~,
\label{120}
\end{equation}
when $q<1/\sqrt{3}$ (i.e.\ $k<k_c=mc/(\sqrt{3}\hbar)$).
The first integral in Eq.~(\ref{118}) can now be written as
\begin{equation}
\frac{1}{2\pi^{2}}\int_0^{k_c} \left(\frac{
\gamma_{\bf k}^2}{1-\gamma_{\bf k}^2}\right)k^2dk=\left(\frac{mc}{\hbar}\right)^3\lambda(\rho)\ ,
\label{121}
\end{equation}
where $\lambda(\rho)$ is given by
\begin{align}
\lambda(\rho)= - \frac{1}{18\pi^2\sqrt{3}}&+\frac{1}{4\pi^2}\int_0^{1/\sqrt{3}}\frac{q^2dq}{\gamma(q,\rho)+1}
\nonumber\\  &-\frac{1}{4\pi^2}\int_0^{1/\sqrt{3}}\frac{q^2dq}{\gamma(q,\rho)-1}\ .
\label{122}
\end{align}
Standard integration methods can be used to find a closed form solution for $\lambda(\rho)$ from Eq.~(\ref{122}). 
A linear approximation in Eq.~(\ref{119}) ($\nu_3=0$ or $E_k=c\hbar k$) gives $\lambda=0.0030$; using the value $\nu_3=-0.1$ quoted above for liquid $^4$He at $T=0$ and $\rho=0.145$ ${\rm g~cm}^{-3}$ gives $\lambda=0.0031$. 

The second integral in Eq.~(\ref{118}) can be calculated in an analytical form when the condition $\delta=\Theta/(mc^2)\ll 1$ with $\Theta=k_B T$ is satisfied. Introducing the new variable $x=c\hbar k/\Theta$ one can in this case write with high accuracy $E_k= c\hbar k$ and $(1+\gamma_{\bf k}^2)/(1-\gamma_{\bf k}^2)= mc/(\hbar k)$. The second integral for $\delta\ll 1$ is then
\begin{equation}
\frac{mc}{2\pi^{2}\hbar}\int_0^{k_c} n_{\bf k}kdk=\frac{m\Theta^2}{2\pi^2c\hbar^{3}}\int_0^\infty \frac{xdx}{e^x-1}=\frac{m\Theta^2}{12c\hbar^3}\ .
\label{123}
\end{equation}
The cutoff can be ignored here, since the integrand is already very small when the cutoff value is reached: $x_c=mc^2/(\sqrt{3}\Theta)\gg 1$ when $\delta\ll 1$.

Combining these two results, at low temperatures (when $\delta=k_B T/(mc^2)\ll 1$) Eq.~(\ref{118}) can be written
\begin{equation}
n_c=n-\left(\frac{mc}{\hbar}\right)^3\lambda(\rho)-\frac{m(k_BT)^2}{12c\hbar^3}\ .
\label{124}
\end{equation}
We note that the last term is important even when $\delta\ll 1$.

When $T=0$ and $\rho =0.145~{\rm g~cm}^{-3}$ the sound velocity is $c=2.37\cdot 10^4$ ${\rm cm/s}$ and  Eq.~(\ref{124}) for the parameter $\lambda=3.00\times10^{-3}$ ($\nu_3=0$) yields the fractions $\rho_c/\rho=0.545$ and $\rho_{ex}/\rho=0.455$. The more accurate parameter $\lambda=3.10\times10^{-3}$ ($\nu_3=-0.1$) yields the fractions $\rho_c/\rho=0.526$ and $\rho_{ex}/\rho=0.474$. 
These values are in good agreement with the prediction from Eq.~(\ref{97}) based on Monte Carlo simulations for liquid $^4$He, namely $\rho_c/\rho=0.528$ and $\rho_{ex}/\rho=0.472$.

Eqs.~(\ref{89}) and (\ref{124}) at $T=0$ lead to the ground state energy per particle as 
\begin{equation}
{\cal E}_0=-\frac{1}{2}\left(mc^2- \frac{m^5c^5\lambda(\rho)}{\rho\hbar^3}\right)\ ,
\label{125}
\end{equation}
where the sound velocity $c$ is given as a function of density by Eq.~(\ref{93}) or Eq.~(\ref{250}).
For example, Eq.~(\ref{125}) at $\rho =0.145~{\rm g~cm}^{-3}$ and $\lambda=0.003$ ($\nu_3=0$) yields for liquid $^4$He the ground state energy ${\cal E}_0/k_B=-7.37~{\rm K}$; with the more accurate value $\lambda=0.0031$ ($\nu_3=-0.1$) the  ground state energy is ${\cal E}_0/k_B=-7.12~{\rm K}$.
The Monte Carlo simulation results for the ground state energy are contained in Eq.~(\ref{92}), which yields 
at $\rho =0.145~{\rm g~cm}^{-3}$ 
the  ground state energy ${\cal E}_0/k_B=-7.15~{\rm K}$. 
This in good agreement with  the theoretical prediction given by Eq.~(\ref{125}), differing by only 0.4\% .

\section{Roton clusters in liquid $^4$He}

In this section we derive the generalized Hartree-Fock (GHF) equation describing roton clusters in liquid $^4$He. A stable cluster consisting of $N$ bound helium atoms can be modelled by the Hamiltonian
\begin{equation}
H=-\sum_{j=1}^N \frac{\hbar^2}{2m}\Delta_j+\sum_{j<k}^N U(|{\bf x}_j-{\bf x}_k|)+\sum_{j=1}^N V({\bf x}_j,t)\ ,
\label{15}
\end{equation}
where $U(|{\bf x}_j-{\bf x}_k|)$ is the two-particle potential between particles in the cluster, and $V({\bf x}_j,t)$ describes the interaction of a particle in the cluster 
with all particles in the bulk at time $t$.
The potential $V({\bf x},t)$ is assumed \cite{Kr,Kru} to be of the form 
\begin{equation}
V({\bf x},t)=\gamma({\bf n},t)(x_1^2+x_2^2+x_3^2)\ ,
\label{16}
\end{equation}
where $\gamma({\bf n},t)$ depends on the unit vector ${\bf n}={\bf x}/|{\bf x}|$ and the time $t$.  
Applying the variational procedure with the trial wavefunction (\ref{2}) to the Hamiltonian (\ref{15}) yields the generalized Hartree-Fock (GHF) time-dependent equation for the one-particle wavefunction $\Psi_N({\bf x},t)$ of the cluster as
\begin{equation}
i\hbar\frac{\partial}{\partial t}\Psi_N({\bf x},t)=\left\{-\frac{\hbar^2}{2m}\Delta+{\cal U}({\bf x},t)\right\}\Psi_N({\bf x},t)\ ,
\label{17}
\end{equation}
with the normalization $\int |\Psi_N({\bf x},t)|^2d{\bf x}=1$. Here ${\cal U}({\bf x},t)$ is the full mean-field potential given by
\begin{equation}
{\cal U}({\bf x},t)=U_{HF}({\bf x},t)+\gamma({\bf n},t)(x_1^2+x_2^2+x_3^2)\ ,
\label{18}
\end{equation}
where $U_{HF}({\bf x},t)$ is the Hartree-Fock potential:
\begin{equation}
U_{HF}({\bf x},t)=(N-1)\int U(|{\bf x}-{\bf x}'|)|\Psi_N({\bf x}',t)|^2d{\bf x}'\ .
\label{19}
\end{equation}
The full force ${\cal F}$ along the direction ${\bf n}=({\rm n}_1,{\rm n}_2,{\rm n}_3)$ at the boundary of the cluster is zero, leading to
\begin{equation}
{\cal F}({\bf n},t)=-\sum_{k=1}^3{\rm n}_k\left(\frac{\partial}{\partial x_k}{\cal U}({\bf x},t)\right)_{{\bf x}=a{\bf n}}=0\ ,
\label{20}
\end{equation}
where the parameter $a$ is the radius of the cluster consisting of $N$ particles. It is assumed that in the stationary state such a cluster has a spherical shape centred at ${\bf x}=0$;  
${\bf x}=a{\bf n}$ is then the stationary boundary point of the cluster in the direction ${\bf n}$. The radius $a$ of the cluster and the components of the unit vector ${\bf n}$ are given by
\begin{equation}
a=\left(\frac{3}{4\pi n}\right)^{1/3}N^{1/3},~~~{\rm n}_k=\frac{x_k}{\sqrt{x_1^2+x_2^2+x_3^2}}\ ,
\label{21}
\end{equation}
where $n$ is the average density of the bulk. Eqs.~(\ref{18}) and (\ref{20}) yield the function $\gamma({\bf n},t)$ in the bulk potential of the form:
\begin{equation}
\gamma({\bf n},t)=\frac{(N-1)}{2a}\int |\Psi_N({\bf x}',t)|^2 F({\bf n},{\bf x}') d{\bf x}'\ ,
\label{22}
\end{equation}
where $F({\bf n},{\bf x}')$ is the force between particles in the cluster,
\begin{equation}
F({\bf n},{\bf x}')=-\sum_{k=1}^3{\rm n}_k\left(\frac{\partial}{\partial x_k}U(|{\bf x}-{\bf x}'|)\right)_{{\bf x}=a{\bf n}}\ .
\label{23}
\end{equation}
Thus the full mean field potential ${\cal U}({\bf x},t)$ in the GHF equation given by Eq.~(\ref{17}) is
\begin{equation}
{\cal U}({\bf x},t)=(N-1)\int {\cal V}({\bf x},{\bf x}')|\Psi_N({\bf x}',t)|^2d{\bf x}'\ ,
\label{24}
\end{equation}
where ${\cal V}({\bf x},{\bf x}')$ is 
\begin{equation}
{\cal V}({\bf x},{\bf x}')= U(|{\bf x}-{\bf x}'|)+(2a)^{-1}F({\bf n},{\bf x}'){\bf x}^2\ . 
\label{25}
\end{equation}

The wavefunction of the GHF equation in the stationary case can be written in the standard form
\begin{equation}
\Psi_N({\bf x},t)=e^{-(i/\hbar){\cal E}_N t}\Phi_N({\bf x})\ . 
\label{26}
\end{equation}
Thus Eq.~(\ref{17}) leads to an eigenenergy ${\cal E}_N$  given by
\begin{equation}
{\cal E}_N=\langle K\rangle_N+(N-1)\langle {\cal V}\rangle_N\ ,
\label{27}
\end{equation}
where the mean field kinetic energy $\langle K\rangle_N$ and the full mean field potential energy $\langle {\cal V}\rangle_N$ are
\begin{equation}
\langle K\rangle_N=\int \Phi_N^{*}({\bf x})\left(-\frac{\hbar^2}{2m}\Delta \right)\Phi_N({\bf x})d{\bf x}\ ,
\label{28}
\end{equation}
\begin{equation}
\langle {\cal V}\rangle_N=\int\int {\cal V}({\bf x},{\bf x}')|\Phi_N({\bf x})|^2|\Phi_N({\bf x}')|^2d{\bf x}d{\bf x}'\ .
\label{29}
\end{equation}

It is worthwhile mentioning that the expectation value $E_N=\langle H\rangle_N$ of the Hamiltonian Eq.~(\ref{15}) with trial wavefunction Eq.~(\ref {2}) can be written in the same form as Eqs.~(\ref{9}) and (\ref{10}) with the replacement of the mean field potential energy $\langle U\rangle_N$ by the full mean field potential energy of the cluster $\langle {\cal V}\rangle_N$ given by Eq.~(\ref {29}).

Eq.~(\ref {27}) shows that the number of particles in the cluster is 
$N=({\cal E}_N+\langle {\cal V}\rangle_N-\langle K\rangle_N)/\langle {\cal V}\rangle_N$ where the eigenenergy ${\cal E}_N$ can be found in the stationary state from Eq.~(\ref{26}).
The evaluation of the cluster number $N$ for a wide range of densities and pressures is given in Appendix D; this also uses the ground state energy obtained in Sec. V. In the particular case when the pressure in the liquid helium is zero this number is given by Eq.~(\ref{204}) as 
\begin{equation}
N=\frac{3\Delta+{\cal E}_0}{\Delta+{\cal E}_0}\ ,
\label{30}
\end{equation}
where $\Delta=-\mu$.
Using the known parameters $\Delta/k_B=8.65~{\rm K}$ and ${\cal E}_0/k_B=-7.15~{\rm K}$ for liquid $^4$He \cite{Kru} at zero pressure, we find that the number of particles in the roton cluster is $N=12.5$.  

The real cluster number $N$ should of course be an integer: either $N=12$ or $N=13$. It is natural to expect that the most stable clusters (with $N\simeq 12.5$ and $P=0$) actually consist of 13 helium atoms, with a central atom surrounding by a shell of 12 atoms situated at the vertices of a regular icosahedron \cite{Kr,Kru}. The stability of this configuration is favored by its having the greatest number (six) of nearest neighbors for each atom in the shell and and also the most compact spherical form. 

A more accurate result for the number of atoms in the roton clusters can be found by numerical simulation of the GHF Eq.~(\ref{17}) with an additional temperature-dependent noise term in the potential ${\cal U}({\bf x},t)$. This stochastic term modeling the collisions of the roton cluster with the surrounding thermalised atoms is important because only stable solutions of the GHF equation should be selected. The numerical simulation of the GHF Eq.~(\ref{17}) with a stochastic source in the potential will be presented elsewhere.

Finally, we show in Fig. 7 the number of atoms in the roton clusters as a function of density $\rho$, given by the approximation Eq.~(\ref{210}) (see Appendix D). Comparison with our previous results \cite{Kr,Kru} suggests that the error of this equation is about $\pm 1$. Fig. 7 indicates that over a the wide range of densities ($0.145~{\rm g~cm}^{-3}<\rho<0.172~{\rm g~cm}^{-3}$) or pressures ($0<P<25~{\rm atm}$) the nearest integer number of atoms in the clusters is $N=13$.  The number of atoms in the cluster can be less than $13$ if the pressure is negative.
 \begin{figure}
\centering{}
  \includegraphics[width=9cm]{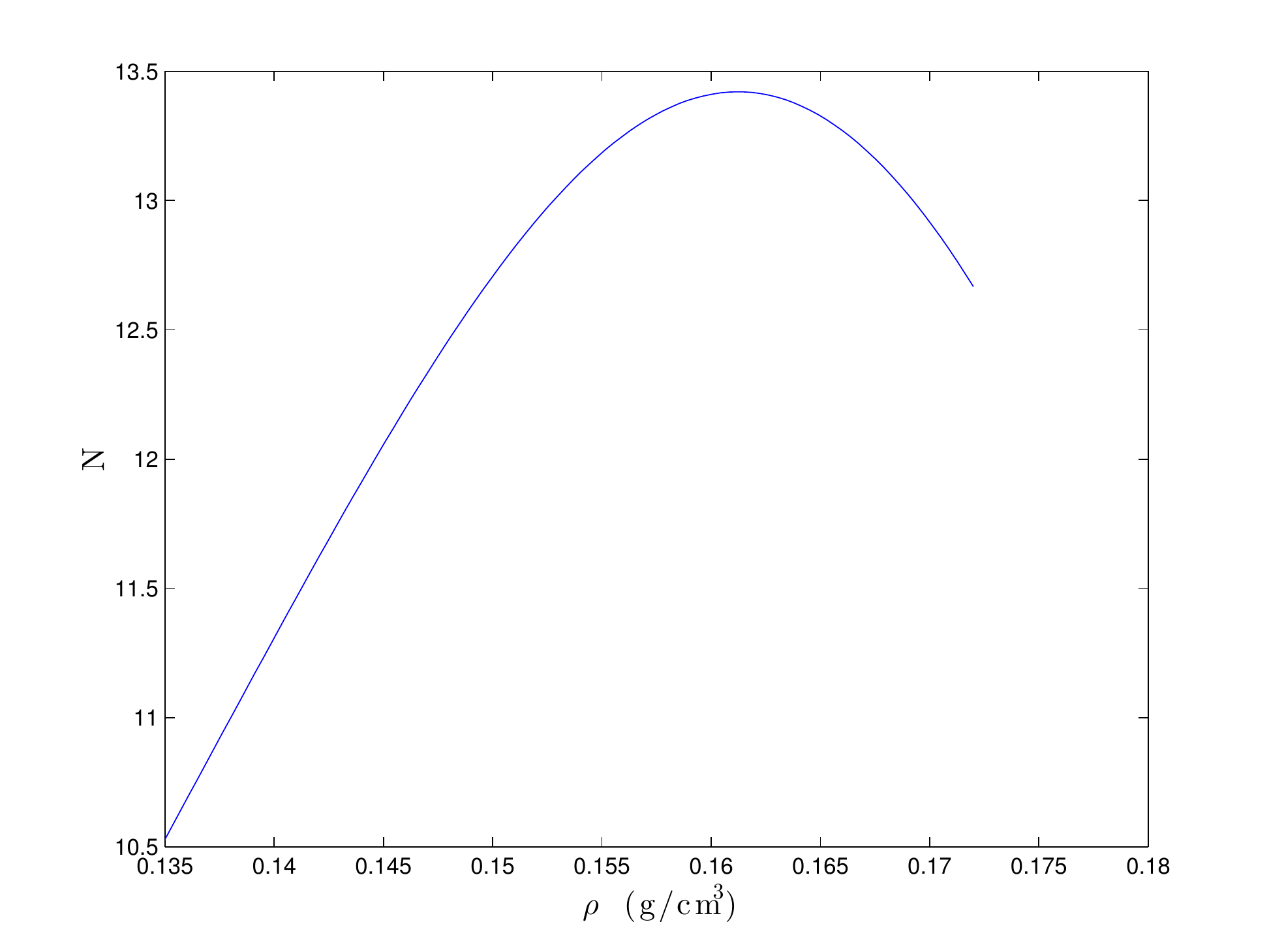}
  \caption{(Color online) Cluster number as a function of density $\rho$, given by Eq.~(\ref{210}). This number can be less than $13$ when the pressure is negative.} 
  \label{F7}
\end{figure}

\section{Conclusions}

We have found a full set of wave equations describing dense Bose fluids, including both nonideal gases and liquid $^4$He. The phonon spectrum and the fraction of condensed particles are calculated for liquid $^4$He at zero temperature for wide range of densities. The theory also allows us to calculate the ground state energy of this quantum liquid in agreement to high accuracy with Monte Carlo simulations and experimental data at low pressure.

It may be instructive to discuss the nearly-ideal regime, when the interaction strength parameter $\sqrt{a_0^3n}$ is small. This limiting case is considered in detail in Appendix C.  
 It is well known that in the Bogoliubov approximation for the hard-sphere model \cite{Hu,Lee,Hua,Leg} the ground state energy per particle to  first-order perturbation in $\sqrt{a_0^3n}$ is
 \begin{equation}
{\cal E}_0^{HS}=\tilde{{\cal E}_0}\left(1+\sigma\sqrt{a_0^3n}\right)\ ,
\label{126}
\end{equation}
where $\tilde{{\cal E}_0}=2\pi a_0\hbar^2n/m$ is the Bogoliubov energy in zeroth-order approximation (i.e.\ the GP energy) and $\sigma=128/(15\sqrt{\pi})$. 
We see that, contrary to the standard variational argument for the ground state energy, this perturbative correction to the GP energy $\tilde{{\cal E}_0}$ leads to a greater ground state energy. Leggett \cite{Legg} has called this result a `pseudo-paradox', 
and has shown that it is \emph{not} due to the replacement of the true interatomic potential by a delta-function pseudo-potential.

In the present theory (Appendix C) the perturbative correction to the GP energy for small $\sqrt{a_0^3n}$ yields in the NIBG approximation the negative value $\sigma=-2\lambda(4\pi)^{3/2}$ (with $\lambda=0.003$). Thus, contrary to the hard-sphere model, our approach based on a realistic interaction potential $U(r)$ leads to 
an estimated ground state energy ${\cal E}_0$ that is less than $\tilde{{\cal E}_0}$, as required by the variational argument. 

The ground state energy is also found in Appendix C (see Eqs.~(\ref{302}) and (\ref{306})) in the more general form   
 \begin{equation}
{\cal E}_0=\tilde{{\cal E}_0}\left(1-\frac{n_{ex}}{n}\right)^2\ ,\quad\frac{n_{ex}}{n}=\varepsilon\left(1-\frac{n_{ex}}{n}\right)^{3/2}\ ,
\label{127}
\end{equation}
where $\varepsilon=\lambda (4\pi)^{3/2}\sqrt{a_0^3n}$. This more general result still satisfies the required inequality, avoiding the `pseudo-paradox'.

The converse inequality ${\cal E}_0^{HS}>\tilde{{\cal E}_0}$ characteristic of the hard-sphere model
should not be understood as a disagreement with our theory. A model with delta-potential intermolecular interaction should be treated by special methods \cite{Hu,Lee,Hua,Leg,Legg} that use the Bogoliubov approximation and renormalization procedures to remove the divergences connected with the singular potential. 
Moreover, as Leggett emphasizes, taking the results from a true hard-sphere model over to the case of a similar but not identical model requires explicit justification
\cite{Hua,Legg}. 

We have also presented in the paper the derivation of a generalized Hartree-Fock equation describing roton clusters in liquid $^4$He at low temperatures. This equation allows us to evaluate the number of bound atoms in each cluster; over a wide range of densities [$0.145~{\rm g~cm}^{-3}<\rho<0.172~{\rm g~cm}^{-3}$] and pressures [$0<P<25~{\rm atm}$] the number of atoms in a roton cluster is $13$.

Finally, we emphasize that the present theory assumes a realistic, and hence nonsingular, interaction potential between atoms, which can contain both repulsive and attractive parts. Furthermore, the theory developed in this paper is consistent in the NIBG and DBG approximations with the variational argument for the ground state energy, both leading to the same results to first order in $\sqrt{a_0^3n}$ (see Appendix C). 
 
\section*{Acknowledgments}

The authors are grateful to Professor H. J. Carmichael for useful discussion of the results of this work.

\appendix

\section{Modified Born approximation}

The standard definition of the s-wave scattering length is given by $1/a_0=-{\rm lim}_{k\rightarrow 0}k\cot\delta_0(k)$, where $\delta_0(k)$ is the phase shift of the s-scattering wave function and $k$ is the wave number. It can also be written as
\begin{equation}
a_0= \lim_{k\rightarrow 0} \frac{m}{\hbar^2}\int_0^{\infty} U(r)\phi_k(r)rdr\ ,
\label{151}
\end{equation}
where $U(r)$ is the scattering potential and $\phi_k(r)$ is an exact wavefunction defined in the scattering theory with appropriate boundary conditions. When the Born approximation is valid the wavefunction $\phi_k(r)$ in Eq.~(\ref{151}) can be replaced by the wavefunction of a free particle in the form $\phi_k^{(0)}(r)=k^{-1}\sin kr $. However, for many scattering potentials $U(r)$ this approximation is meaningless because  the integral in Eq.~(\ref{151}) diverges. 

In the modified Born approximation (MBA) \cite{Kru} the wavefunction in Eq.~(\ref{151}) is instead approximated by
\begin{equation}
\phi_k(r)=\theta(r-a_0) k^{-1}\sin kr\ , 
\label{152}
\end{equation}
where $\theta(r)$ is the Heaviside unit step function. Thus the wavefunction in Eq.~(\ref{152}) is zero for $r<a_0$; the region $r<a_0$ is unattainable for slow particles ($k\rightarrow 0$) because the cross-section for s-scattering waves is $\sigma_s=4\pi a_0^2$.

Eqs.~(\ref{151}) and (\ref{152}) lead to an equation for the s-wave scattering length \cite{Kru}:
\begin{equation}
a_0= \frac{m}{\hbar^2}\int_{a_0}^{\infty} U(r)r^2dr=\frac{m}{\hbar^2}\int_{0}^{\infty} \tilde{U}(r)r^2dr\ ,
\label{153}
\end{equation}
where the effective potential $\tilde{U}(r)$ incorporates the cut-off: $\tilde{U}(r)=0$ for $r<a_0$ and $\tilde{U}(r)=U(r)$ for $r\geq a_0$. 

In the case of gas or liquid $^4$He the intermolecular interactions are given by the Lennard-Jones potential
\begin{equation}
U(r)=4\epsilon\left[\left(\frac{r_0}{r}\right)^{12}-\left(\frac{r_0}{r}\right)^{6}\right]\ .
\label{154}
\end{equation}
The minimum of the potential occurs at $r_m=2^{1/6}r_0$. 
To good accuracy, the parameters of the Lennard-Jones potential for $^4$He are $\epsilon/k_B=10.6~{\rm K}$ and $r_m=2.98$~\AA. 

Eqs.~(\ref{153}) and (\ref{154}) together yield a fifth-order \cite{Kru} algebraic equation
\begin{equation}
\Lambda^5-3\Lambda^2-\Lambda_0=0\ ,
\label{155}
\end{equation}
where the parameters $\Lambda$ and $\Lambda_0$ are
\begin{equation}
\Lambda=\left(\frac{r_0}{a_0}\right)^2,~~~\Lambda_0=\frac{9\hbar^2}{4\epsilon mr_0^2}\ .
\label{156}
\end{equation}
The solution of Eq.~(\ref{155}) with the parameters given above for the Lennard-Jones potential in $^4$He leads to an s-wave scattering length $a_0=2.20$~\AA.

\section{Functions of state in liquid $^4$He}

We present here analytical approximations for the sound velocity, pressure, ground state energy and chemical potential in liquid $^4$${\rm He}$ at zero temperature. This approach is based on the observation that the sound velocity in the liquid $^4$${\rm He}$ at zero temperature is to high accuracy a linear function of the density $\rho$. Hence we can write
\begin{equation}
c=\sigma_0+\sigma_1\rho\ ,
\label{250}
\end{equation}
where the parameters $\sigma_0=-4.116\cdot 10^4$ and $\sigma_1=4.473\cdot 10^4$ (in ${\rm c.g.s.}$ units) may be found from experimental and numerical data. 
For comparison, while the form of the sound velocity in Eq.~(\ref{93}) is not exactly linear, it is close 
to it with a high accuracy (see Fig. 4). 
We note that the sound velocity given by Eq.~(\ref{250}) has a physical and mathematical sense because it is true linear function of the density in the liquid $^4$${\rm He}$ at zero temperature. 

Integrating $\partial P/\partial \rho=c^2$, where the sound velocity is given by Eq.~(\ref{250}), we have for the pressure
\begin{equation}
P=b_0+b_1\rho+b_2\rho^2+b_3\rho^3\ ,
\label{251}
\end{equation}
with $b_1= \sigma_0^2$, $b_2= \sigma_0\sigma_1$ and $b_3= \sigma_1^2/3$; the integration constant $b_0$ can be found from any boundary condition of the form $P=P_0$ at $\rho =\rho_0$. In particular, we may take $\rho_0=0.145$ ${\rm g ~cm}^{-3}$ and $P_0=0$. The resulting parameters in ${\rm c.g.s.}$ units are: $b_0=-6.19\cdot 10^7$, $b_1=1.694\cdot 10^9$, $b_2=-1.841\cdot 10^{10}$, and $b_3=6.669\cdot 10^{10}$.

The ground state energy per particle satisfies $\partial {\cal E}_0/\partial \rho=mP/\rho^2$, where the pressure is given by Eq.~(\ref{251}). Thus we have ${\cal E}_0$ as a function of $\rho$ in the form
\begin{equation}
{\cal E}_0=-\frac{a_0}{\rho}+a+a_2\rho+\frac{1}{2}a_3\rho^2+a_1{\rm ln}\frac{\rho}{\rho_0}\ ,
\label{252}
\end{equation}
where the parameters $a_k$ ($k=0,1,2,3$) are just $a_k=mb_k$. The integration constant $a$ in this equation can again be found from a known value: in this case we know that the ground state energy is ${\cal E}_0/k_B=-7.15~{\rm K}$ at $\rho=0.145$ ${\rm g ~cm}^{-3}$. This yields the value $a=mb$ where $b=1.394\cdot 10^9$ in ${\rm c.g.s.}$ units. 

The chemical potential at zero temperature can be found from $\mu={\cal E}_0+mP/\rho$, leading to
\begin{equation}
\mu=a+a_1+2a_2\rho+\frac{3}{2}a_3\rho^2+a_1{\rm ln}\frac{\rho}{\rho_0}\ .
\label{253}
\end{equation}

Using the relations given by Eqs.~(\ref{88}) and (\ref{89}) we can now write the coupling parameters $G$ and $g$ as the analytical functions of density 
\begin{equation}
G=-\frac{m^3c^4}{2\rho{\cal E}_0}\ ,~~~g=-\frac{m^2c^2}{\rho}\left(1+\frac{mc^2}{2{\cal E}_0}\right)\ .
\label{254}
\end{equation}
The condensate and excited fractions now follow from Eq.~(\ref{89}) as
\begin{equation}
\frac{\rho_c}{\rho}=-\frac{2{\cal E}_0}{mc^2}\ ,~~~\frac{\rho_{ex}}{\rho}=1+\frac{2{\cal E}_0}{mc^2}\ .
\label{255}
\end{equation}
The sound velocity $c$ and the ground state energy ${\cal E}_0$ in Eqs.~(\ref{254}) and (\ref{255}) are given by Eqs.~(\ref{250}) and (\ref{252}).

The coupling parameters $G$ and $g$ given by Eqs.~(\ref{254}), (\ref{250}) and (\ref{252})  coincide to high accuracy with the curves in Eqs.~(\ref{95}) and (\ref{96}) presented in Fig. 5. The fractions given by Eqs.~(\ref{255}), (\ref{250}) and (\ref{252}) also coincide to high accuracy with the curves demonstrated in Fig. 6. 

Thus, we have for a wide range of densities an alternative analytical representation for the sound velocity, pressure, ground state energy and chemical potential in the liquid $^4$${\rm He}$ at zero temperature.

\section{NIBG and DBG approximations}

In this Appendix we consider perturbative corrections in the nearly ideal and dilute Bose gase theories, with the small parameter $\sqrt{a_0^3n}$. 
Our approach includes a cut-off in momentum space at the wave number $k_c=mc_0/\hbar$, where $c_0$ is the second sound. In the case of the NIBG only the first term in Eq.~(\ref{119}) should be taken into account, which yields the linear phonon spectrum $E_k=c\hbar k$ for $k<mc_0/\hbar$. 
In this approximation  Eq.~(\ref{73}) gives the sound velocity as
\begin{equation}
c=\frac{\hbar}{m}\sqrt{4\pi a_0n_c}\ .
\label{301}
\end{equation}
For consistency, it is important that the condition $\sqrt{a_0^3n}\ll 1$ and the cut-off procedure together imply the necessary condition $a_0k\ll 1$ of the NIBG theory. 
At $T=0$, Eqs.~(\ref{124}) and (\ref{301}) lead to
\begin{equation}
(n-n_c)^2=\lambda^2(4\pi a_0n_c)^3\ ,
\label{302}
\end{equation}
with $\lambda=0.003$. 
To first  order in the small parameter $\sqrt{a_0^3n}$,  
Eqs.~(\ref{301}) and (\ref{302}) give the condensate fraction $n_c/n$ and the sound velocity $c$ as
\begin{equation}
\frac{n_c}{n}=1-\Gamma\sqrt{a_0^3n}\ ,\quad c=\frac{\hbar}{m}\sqrt{4\pi a_0n}\left(1-\frac{1}{2}\Gamma\sqrt{a_0^3n}\right)\ ,
\label{303}
\end{equation}
where 
\begin{equation}
\Gamma=\lambda(4\pi)^{3/2}\ .
\end{equation}
Hence for $\lambda=0.003$ we have $\Gamma=0.134$. In contrast, in the Bogoliubov theory for the hard-sphere model, the fraction $n_c/n$ is calculated without a cut-off in  momentum space, leading to the value $\Gamma=8/(3\sqrt{\pi})=1.50$. 
The  cut-off procedure in our approach automatically keeps the necessary condition $a_0k\ll 1$. 

The ground state energy per particle can be found in the first perturbation order from Eqs.~(\ref{71}) and (\ref{303}) as
\begin{equation}
{\cal E}_0=\frac{mc^2n_c}{2n}=\frac{2\pi a_0\hbar^2n}{m}\left(1-2\Gamma\sqrt{a_0^3n}\right)\ ,
\label{304}
\end{equation}
and the chemical potential fram Eqs.~(\ref{81}) and (\ref{304}) as
\begin{equation}
\mu=\frac{4\pi a_0\hbar^2n}{m}\left(1-\frac{5}{2}\Gamma\sqrt{a_0^3n}\right)\ .
\label{305}
\end{equation}

Using the NIBG theory with a hard-sphere pseudo-potential \cite{Hu,Lee,Hua,Leg} still gives results of the form (\ref{304}) and (\ref{305}),
but with yet another value for $\Gamma$, $\Gamma=-64/(15\sqrt{\pi})=-2.41$. the negative value of $\Gamma$ is connected with the `pseudo-paradox' for the ground state energy in the hard-sphere model discussed in Sec. VIII. 

Alternatively, in the NIBG approximation one can find the fraction $n_c/n$ directly from the cubic Eq.~(\ref{302}) and then the sound velocity $c$ from Eq.~(\ref{301}). In this approach the ground energy ${\cal E}_0$ is given by Eq.~(\ref{74}) as
\begin{equation}
{\cal E}_0=\frac{2\pi a_0\hbar^2n_c^2}{mn}=\frac{2\pi a_0\hbar^2n}{m}\left(1-\frac{n_{ex}}{n}\right)^2\ ,
\label{306}
\end{equation}
which is consistent to first perturbation order with (\ref{304}).

For the DBG approximation, Eq.~(\ref{124}) at $T=0$ yields the excitation fraction in the form
\begin{equation}
\frac{n_{ex}}{n}=\frac{\lambda}{n}\left(\frac{mc}{\hbar}\right)^3,~~~c=\sqrt{\frac{g(n)n_c}{m}}\ ,
\label{307}
\end{equation}
again with $\lambda=0.003$. The ground state energy for this approximation is given by Eq.~(\ref{71}). 
The coupling parameter $g(n)$ may be expanded in the small dimensionless density parameter $\sqrt{a_0^3n}$ as
\begin{equation}
g(n)=g_0+g_1\sqrt{a_0^3n}+... \ ,\quad g_0= \frac{4\pi a_0\hbar^2}{m}\ ,
\label{308}
\end{equation}
where $g_1$ is a positive constant (since $a_s(n)<a_0$ in the DBG approximation). 

Using  Eqs.~(\ref{307}), (\ref{308}) and (\ref{71}) one can show that the DBG theory leads to additional corrections for the ground state energy given by Eq.~(\ref{304}) only in the second order in the small parameter $\sqrt{a_0^3n}$.
To first order in this parameter, the NIBG and DBG approximations both lead to the same results for the sound velocity, condensate and excitation fractions, the chemical potential and the ground state energy.  

\section{Cluster numbers in liquid $^4$He}

When $N\gg 1$ the average kinetic energy can be evaluated as $\langle K\rangle_N\simeq -\langle {\cal V}\rangle_N$ where $\langle {\cal V}\rangle_N<0$. Hence it follows from Eq.~(\ref{27}) that the number of particles in the roton cluster is 
\begin{equation}
N=\frac{{\cal E}_N+2\langle {\cal V}\rangle_N}{\langle {\cal V}\rangle_N}\ .
\label{201}
\end{equation}
Here $\langle K\rangle_N$ and $\langle {\cal V}\rangle_N$ are slowly varying functions of $N$ when $N\gg 1$.
The chemical potential of the roton cluster for $N\gg 1$ is 
\begin{equation}
\mu={\cal E}_N-{\cal E}_0\ ,
\label{202}
\end{equation}
where ${\cal E}_0$ is the average energy of the helium atoms in the bulk. This energy is given by ${\cal E}_0=\lim_{N\rightarrow \infty} E_0(N)/N$ where $E_0(N)$ is the ground state energy of the $N$ body quantum system. 

The full mean-field potential energy of the cluster $\langle {\cal V}\rangle_N$ given by Eq.~({\ref 123}) can be approximated for $N\gg 1$ by
\begin{equation}
\langle {\cal V}\rangle_N\simeq \mu-{\cal E}_0\ .
\label{203}
\end{equation}
Thus it follows from Eqs.~(\ref{201}),  (\ref{202}) and (\ref{203}) that the number of particles in the roton cluster is
\begin{equation}
N=\frac{3\mu-{\cal E}_0}{\mu-{\cal E}_0}\ ,
\label{204}
\end{equation}
when the condition $N\gg 1$ is satisfied.

The chemical potential of the roton cluster \cite{Kru} can also be written as
\begin{equation}
\mu=U(r_m)-U(2q_0)\ ,\quad q_0=\left(\frac{3}{4\pi n}\right)^{1/3}\ .
\label{205}
\end{equation}
Here $U(r_m)=-\epsilon$ is the minimum of the potential energy in the Lennard-Jones potential, and $2q_0$ is the average distance between atoms in liquid helium. For example, the mass density $\rho=mn=0.145~{\rm g~cm}^{-3}$ yields $2q_0=4.44$~\AA, which is close to the value $2a_0=4.4$~\AA. 
The second term in (\ref{205}) can be rewritten as a explicit function of the density,
\begin{align}
\epsilon_0(n)&=-U(2q_0)\nonumber\\
&=-4\epsilon\left[\left(\frac{r_0}{2q_0}\right)^{12}-\left(\frac{r_0}{2q_0}\right)^{6}\right]
\label{206}\\
&=4\epsilon (v_0^2n^2-v_0^4n^4)\ ,\quad v_0=\frac{\pi r_0^3}{6}\ .
\label{207}
\end{align}
The roton gap $\Delta$ as a function of density $n$ is then \cite{Kru}:
\begin{equation}
\Delta=-\mu,~~~\mu=\epsilon_0(n)-\epsilon\ .
\label{208}
\end{equation}
Eqs.~(\ref{207}) and (\ref{208}) yield the gap $\Delta(\rho)$ as a function of the mass density $\rho=mn$ in the form
\begin{equation}
\Delta(\rho)=\epsilon -\kappa_2\rho^2+\kappa_4\rho^4,
\label{209}
\end{equation}
where $\kappa_2=4\epsilon/\rho_0^2$, $\kappa_4=4\epsilon/\rho_0^4$  and  $\rho_0=6m/(\pi r_0^3)$. For a mass density $\rho=0.145~{\rm g~cm}^{-3}$ this gives the value $\Delta/k_B=8.73~{\rm K}$, which is in a good agreement with the experimental value $\Delta/k_B=8.65~{\rm K}$. 

Finally,  Eqs.~(\ref{204}),  (\ref{209}) and (\ref{92}) lead to an approximate equation for the cluster number $N$ as a function of density,  
\begin{align}
N&= 1 + 2\frac\Delta{\Delta+{\cal E}_0}
\label{210}\\
&=\frac{3\epsilon+\alpha_1\rho+(\alpha_2/2-3\kappa_2)\rho^2+(\alpha_3/3)\rho^3+3\kappa_4\rho^4}{\epsilon+\alpha_1\rho+(\alpha_2/2-\kappa_2)\rho^2+(\alpha_3/3)\rho^3+\kappa_4\rho^4}\nonumber\ .
\end{align}
We show in Fig.~7 the number of atoms in the roton clusters for liquid $^4$${\rm He}$ predicted by this equation.

At low temperatures and typical pressures the stable clusters in liquid helium  consist of $13$ bound helium atoms \cite{Kr,Kru}, presumably in the form of a central atom surrounded by an icosahedral shell of $12$ atoms. 
The stability of this configuration favor by its having the greatest number (six) of nearest neighbors for each atom in a shell and this configuration has the most compact spherical form as well.
Fig. 7 indicates that over a the wide range of densities ($0.145~{\rm g~cm}^{-3}<\rho<0.172~{\rm g~cm}^{-3}$) or pressures ($0<P<25~{\rm atm}$) the nearest integer number of atoms in the clusters is $N=13$.  The number of atoms in the cluster can be less than $13$ if the pressure is negative.


\end{document}